\documentclass{ectj}

\usepackage{amsfonts,amssymb,graphics,epsfig,verbatim,bm,latexsym,amsmath,url,amsbsy}
\usepackage{amsfonts,bbm,mathrsfs,color,enumitem,graphicx,subcaption,natbib}

\pdfoutput=1

\newtheorem{theorem}{Theorem}
\newtheorem{assumption}{Assumption}

\renewcommand{\thesection}{\arabic{section}}
\renewcommand{\theequation}{\arabic{section}.\arabic{equation}}

\received{...}
\accepted{...}
\volume{00}

\DeclareMathOperator*{\argmin}{arg\,min}

\newcommand{\E}{\mathbb{E}}

\newcommand{\V}{\mathbb{V}}
\renewcommand{\P}{\mathbb{P}}

\newcommand{\I}{\mathbbm{1}}

\newcommand{\US}{\mathtt{US}}
\newcommand{\RBC}{\mathtt{RBC}}
\newcommand{\BC}{\mathtt{BC}}
\newcommand{\MSE}{\mathtt{MSE}}

\newcommand{\irbc}{I_\RBC}

\newcommand{\h}{h}
\renewcommand{\b}{b}

\newcommand{\bbeta}{\boldsymbol{\beta}}

\newcommand{\be}{\mathbf{e}}

\newcommand{\br}{\mathbf{r}}

\newcommand{\TR}{\mathtt{TO}}

\newcommand{\x}{c}

\title[Robust Bias Corrected Inference in RD Designs]{Optimal Bandwidth Choice for Robust Bias Corrected Inference in Regression Discontinuity Designs\footnote{A preliminary version of this paper circulated under the title ``Coverage Error Optimal Confidence Intervals for Regression Discontinuity Designs'' (first draft: June 26, 2016). We thank Josh Angrist, Zhuan Pei, Rocio Titiunik, and Gonzalo Vazquez-Bare for comments. Cattaneo gratefully acknowledges financial support from the National Science Foundation (SES 1357561 and SES 1459931). Farrell gratefully acknowledges financial support from the Richard N. Rosett and John E. Jeuck Fellowships.}}

\author[Calonico, Cattaneo, and Farrell]{Sebastian~Calonico$^{\dagger}$,
                Matias~D.~Cattaneo$^{\ddagger}$ and Max~H.~Farrell$^{\dagger\dagger}$}

\address{$^{\dagger}$Department of Health Policy and Management, Columbia University, USA.}
\email{sebastian.calonico@columbia.edu}

\address{$^{\ddagger}$Department of Operations Research and Financial Engineering, Princeton University,USA.}
\email{cattaneo@princeton.edu}

\address{$^{\dagger\dagger}$Booth School of Business, University of Chicago, USA.}
\email{max.farrell@chicagobooth.edu}

\def\AmSTeX{$\cal A$\kern-.1667em\lower.5ex\hbox{$\cal M$}\kern-.125em
    $\cal S$-\TeX}
\def\BibTeX{{\rm B\kern-.05em{\sc i\kern-.025em b}\kern-.08em
    T\kern-.1667em\lower.7ex\hbox{E}\kern-.125emX}}

\begin{document}

    \begin{abstract}

        Modern empirical work in Regression Discontinuity (RD) designs often employs local polynomial estimation and inference with a mean square error (MSE) optimal bandwidth choice. This bandwidth yields an MSE-optimal RD treatment effect estimator, but is by construction invalid for inference. Robust bias corrected (RBC) inference methods are valid when using the MSE-optimal bandwidth, but we show they yield suboptimal confidence intervals in terms of coverage error. We establish valid coverage error expansions for RBC confidence interval estimators and use these results to propose new inference-optimal bandwidth choices for forming these intervals. We find that the standard MSE-optimal bandwidth for the RD point estimator is too large when the goal is to construct RBC confidence intervals with the smallest coverage error. We further optimize the constant terms behind the coverage error to derive new optimal choices for the auxiliary bandwidth required for RBC inference. Our expansions also establish that RBC inference yields higher-order refinements (relative to traditional undersmoothing) in the context of RD designs. Our main results cover sharp and sharp kink RD designs under conditional heteroskedasticity, and we discuss extensions to fuzzy and other RD designs, clustered sampling, and pre-intervention covariates adjustments. The theoretical findings are illustrated with a Monte Carlo experiment and an empirical application, and the main methodological results are available in \texttt{R} and \texttt{Stata} packages. 


    \end{abstract}


\section{Introduction}
\label{sec:intro}

The Regression Discontinuity (RD) design is widely used in program evaluation, causal inference, and treatment effect settings. (For general background on these settings, see \citet*{Imbens-Rubin_2015_Book} and \citet*{Abadie-Cattaneo_2018_ARE}, and references therein.) In recent years, RD has become one of the prime research designs for the analysis and interpretation of observational studies in social, behavioral, biomedical, and statistical sciences. For introductions to RD designs, literature reviews, and background references, see \citet*{Imbens-Lemieux_2008_JoE}, \citet*{Lee-Lemieux_2010_JEL}, \citet*{Cattaneo-Escanciano_2017_AIE}, and \citet*{Cattaneo-Idrobo-Titiunik_2019_Book,Cattaneo-Idrobo-Titiunik_2020_Book}.

Modern empirical work in RD designs often employs a mean square error (MSE) optimal bandwidth for local polynomial estimation of and inference on treatment effects.\footnote{See \citet*{Imbens-Kalyanaraman_2012_REStud}, \citet*{Calonico-Cattaneo-Titiunik_2014_ECMA}, \citet*{Arai-Ichimura_2016_EL,Arai-Ichimura_2018_QE}, \citet*{Calonico-Cattaneo-Farrell-Titiunik_2019_RESTAT}, and references therein. \citet*{Cattaneo-VazquezBare_2016_ObsStud} gives a general discussion of bandwidth/neighborhood selection methods in RD designs.} This MSE-optimal bandwidth choice yields an MSE-optimal RD point estimator, but is by construction invalid for inference. Robust bias corrected (RBC) inference methods provide a natural solution to this problem: RBC confidence intervals and related inference procedures remain valid even when the MSE-optimal bandwidth is used (\citealp*{Calonico-Cattaneo-Titiunik_2014_ECMA,Calonico-Cattaneo-Farrell-Titiunik_2019_RESTAT}). In this paper, we show that this choice of bandwidth is suboptimal when the goal is to construct RBC confidence intervals with minimal coverage error (CE), and we establish a new bandwidth choice delivering CE-optimal RBC confidence interval estimators or, analogously, minimizing the error in rejection probability of the associated hypothesis testing procedures for RD treatment effects.

Our main results are valid coverage error expansions for local polynomial RBC confidence interval estimators. The precise characterization offered by these expansions allows us to study bandwidth selection in detail, and to propose several novel bandwidth choices that are optimal for inference. First and foremost, we derive a CE-optimal bandwidth choice designed to minimize coverage error of the \emph{interval} estimator, which is a fundamentally different goal than minimizing mean square error of the \emph{point} estimator. The MSE- and CE-optimal bandwidths are therefore complementary, as both can be used in empirical work to construct, respectively, optimal point estimators and optimal inference procedures for RD treatment effects. For example, we find that in the case of the popular local linear RD estimator, if the sample size is $n=500$, then shrinking the MSE-optimal bandwidth by approximately $27\%$ leads to RBC confidence intervals with the fastest coverage error decay rate. Further, we use our expansions to derive bandwidth choices that trade off coverage error against interval length, which is conceptually analogous to trading size and power of the associated statistical tests, while retaining asymptotically correct coverage (or size control). Finally, by examining the leading constant terms of our coverage error expansions, we can deliver novel optimal choices for the auxiliary bandwidth required for RBC inference. We also provide plug-in, data-driven bandwidth selectors for use in practice and illustrate their performance with real and simulated data. 

Our theoretical results prove that RBC confidence interval estimators have coverage error strictly smaller (i.e., vanishing faster) than those of interval estimators based on undersmoothing, as long as enough smoothness of the underlying conditional expectation functions is available to at least characterize the MSE of the RD point estimator, the most natural case in empirical applications. RBC intervals are as good as their undersmoothed counterparts when no additional smoothness is available beyond what is needed to quantify the asymptotic bias of the t-test statistic. These results, coupled with our bandwidth selectors, provide precise theory-based guidance for empirical practice employing RD designs: RBC confidence interval estimators constructed with the CE-optimal, and even with the MSE-optimal, bandwidth choice dominate the alternative procedures in terms of coverage error performance.

Our main theoretical results focus on sharp RD designs with heteroskedastic data, covering both levels (standard sharp RD design) as well as derivatives (kink and higher-order RD designs). The latter case being of interest in, for example, \citet*{Card-Lee-Pei-Weber_2015_ECMA,Card-Lee-Pei-Weber_2017_AIE}, \citet*{Dong-Lewbel_2015_ReStat}, \citet*{Cerulli-Dong-Lewbel-Poulsen_2017_AIE}, and \citet*{Ganong-Jager_2018_JASA}. We also discuss extensions to fuzzy, geographic, multi-score, and multi-cutoff RD designs (\citealp*{Hahn-Todd-vanderKlaauw_2001_ECMA,Papay-Willett-Murnane_2011_JoE,Keele-Titiunik_2015_PA,Cattaneo-Keele-Titiunik-VazquezBare_2016_JOP}), as well as to clustered data and/or inclusion of pre-intervention covariates (\citealp*{Lee-Card_2008_JoE,Bartalotti-Brummet_2017_AIE,Calonico-Cattaneo-Farrell-Titiunik_2019_RESTAT}). Our results can also be applied to other RD methods and settings such as those considered in \citet*{Xu_2017_JoE}, \citet*{Dong_2019_JBES}, \citet*{Dong-Lee-Gou_2019_wp}, \citet*{Chiang-Hsu-Sasaki_2019_JoE}, and \citet*{He-Bartalotti_2019_wild}.

Finally, we remark that our discussion of inference-optimal bandwidth selection, as well as all treatments of MSE-optimal choices, are within the context of local polynomial methods (\citealp*{Fan-Gijbels_1996_Book}) under continuity assumptions of the underlying conditional expectation functions. CE- and MSE-optimal bandwidth choices should not be used when the goal is to employ local randomization assumptions in the context of RD designs (\citealp*{Cattaneo-Frandsen-Titiunik_2015_JCI}), because in this setting the underlying assumptions are different and the targeted neighborhood around the cutoff is conceptually distinct. As such, the appropriate neighborhood under local randomization cannot be generated by MSE- or CE-optimal bandwidth choices, and other methods are more appropriate: see Section 3 in \citet*{Cattaneo-Frandsen-Titiunik_2015_JCI} for one example. For further discussion of these different assumptions and methodologies, as well as comparisons between neighborhood selectors, see \citet*{Cattaneo-VazquezBare_2016_ObsStud}, \citet*{Cattaneo-Titiunik-VazquezBare_2017_JPAM}, and \citet*{Sekhon-Titiunik_2017_AIE}.

The rest of the paper proceeds as follows. Section \ref{sec:setup} presents the RD setup and gives a brief, but self-contained, introduction to standard estimation and inference methods. Section \ref{sec:mainresults} gives the main results of the paper: valid higher-order coverage error expansions for commonly used confidence intervals as well as CE-optimal and related bandwidth choices. Section \ref{sec:implementation} discusses implementation and other practical issues. Section \ref{sec:extensions} briefly outlines several extensions, while numerical results using real and simulated data are reported in Section \ref{sec:numerical}. Finally, Section \ref{sec:conclusion} concludes. The supplemental appendix (SA, hereafter) contains all technical details and proofs, as well as more discussion of methodological, implementation, and numerical issues. \citet*{Calonico-Cattaneo-Farrell-Titiunik_2017_Stata} details general purpose \texttt{Stata} and \texttt{R} software packages implementing our main methodological results.

\section{Setup}\label{sec:setup}

We assume the researcher observes a random sample $(Y_i,T_i,X_i)'$, $i=1,2,\dots,n$, where $Y_i$ denotes the outcome variable of interest, $T_i$ denotes treatment status, and $X_i$ denotes an observed continuous score or running random variable, which determines treatment assignment for each unit in the sample. In the canonical sharp RD design, all units with $X_i$ not smaller than a known threshold $\x$ are assigned to the treatment group and take up treatment, while all units with $X_i$ smaller than $\x$ are assigned to the control group and do not take-up treatment, so that $T_i=\I(X_i\geq\x)$. Using the potential outcomes framework, $Y_i=Y_i(0)\cdot(1-T_i)+Y_i(1)\cdot T_i$, with $Y_i(1)$ and $Y_i(0)$ denoting the potential outcomes with and without treatment, respectively, for each unit. The parameters of interest in sharp RD designs are either the average treatment effect at the cutoff or its derivatives:
\[\tau_\nu = \tau_\nu(\x) 
= \left.\frac{\partial^\nu}{\partial x^\nu}\E[Y_i(1)-Y_i(0)|X_i=x]\right|_{x=\x}.\]
where here and elsewhere we drop evaluation points of functions when it causes no confusion. With this notation, $\tau_0$ corresponds to the standard sharp RD estimand, while $\tau_1$ denotes the sharp kink RD estimand (up to scale). In Section \ref{sec:extensions}, we discuss imperfect treatment compliance (i.e., fuzzy RD designs) and other extensions of this basic RD setup. Identification of $\tau_\nu$, as well as estimation and inference using local polynomial regression methods, proceed under the following standard regularity conditions. 

\begin{assumption}[RD]\label{ass:srd}
	For all $x\in[x_l,x_u]$, where $x_l<\x<x_u$, and $t\in\{0,1\}$: $\E[Y_i(t)|X_i=x]$ is $S\geq \min\{1,\nu\}$ times continuously differentiable with an $S^{\text{th}}$ derivative that is H\"older continuous with exponent $a\in(0,1]$; the Lebesgue density of $X_i$, $f(x)$, and $\V[Y_i(t)|X_i=x]$ are positive and continuous; $\E[|Y_i(t)|^\delta|X_i=x]$, $\delta > 8$, is continuous; the Lebesgue density of $(Y(t), X)$, $f_{y_t x}(\cdot)$, is positive and continuous.
\end{assumption}

``Flexible'' (i.e., nonparametric) local polynomial least squares estimators are the standard approach for estimation and inference in RD designs. The idea is to first choose a neighborhood around the cutoff $\x$ via a positive bandwidth choice $\h$, and then employ (local) weighted polynomial regression using only observations with score $X_i$ laying within the selected neighborhood. That is, 
\[\hat{\tau}_{\nu}(h) = \nu!\be_\nu'\hat{\bbeta}_{+,p}(h) - \nu!\be_\nu'\hat{\bbeta}_{-,p}(h), \qquad \nu=0,1,2,\dots,p,\]
where $\be_\nu$ denotes the conformable $(\nu+1)$-th unit vector, and $\hat{\bbeta}_{-,p}(h)$ and $\hat{\bbeta}_{+,p}(h)$ correspond to the weighted least squares coefficients given by
\begin{align*}
	\hat{\bbeta}_{-,p}(h)
	&= \argmin_{\bbeta\in\mathbb{R}^{p+1}} \sum_{i=1}^n \I(\x>X_i) \; \big(Y_i - \br_{p}(X_i-\x)'\bbeta\big)^2 \; K_h(X_i-\x),\\
	\hat{\bbeta}_{+,p}(h)
	&= \argmin_{\bbeta\in\mathbb{R}^{p+1}} \sum_{i=1}^n \I(\x\leq X_i) \; \big(Y_i - \br_{p}(X_i-\x)'\bbeta\big)^2 \; K_h(X_i-\x),
\end{align*}
with $\br_{p}(x)=(1,x,\cdots,x^p)'$ and $K_h(\cdot)=K(\cdot/h)/h$ for a kernel (weighting) function $K(\cdot)$. The kernel is assumed to obey the following regularity conditions.

\begin{assumption}[Kernel]\label{ass:kernel}
	$K(u)=\I(u<0)k(-u)+\I(u\geq 0)k(u)$, where $k(\cdot):[0,1]\mapsto\mathbb{R}$ is bounded and continuous on its support, positive $(0,1)$, zero outside its support, and either is constant or $(1, K(u) \br_{3(p+1)}(u)')$ is linearly independent on $(-1,1)$.
\end{assumption} 

The kernel and bandwidth serve to localize the regression fit near the cutoff. The choice of bandwidth, $\h$, is the key parameter when implementing the RD estimator, and we discuss this choice in detail below. The most popular choices of kernel are the uniform kernel and the triangular kernel, which give equal weighting and linear down-weighting to the observations with $X_i\in[\x-h,\x+h]$, respectively. Finally, although our results cover any choice of $p\geq0$, the preferred choice of polynomial order for point estimation is $p=1$ (i.e., local-linear RD treatment effect estimator) because of the poor behavior of higher-order polynomial approximations at or near boundary points. See Section 2.1.1 of \citet*{Calonico-Cattaneo-Titiunik_2015_JASA} and \cite*{Gelman-Imbens_2019_JBES} for more discussion.

\subsection{MSE-Optimal Bandwidth Choice and Point Estimation}

Selecting the bandwidth $h$ or, equivalently, the neighborhood around the cutoff $\x$, is challenging in applications. The default approach in modern empirical work is to minimize an approximation to the MSE of the point estimator $\hat{\tau}_{\nu}(h)$, or some other closely related quantity. Under standard regularity conditions, the conditional MSE of $\hat{\tau}_{\nu}(h)$ can be approximated as $\h\to0$ and $nh\to\infty$ as follows:
\begin{equation}\label{eq:mse}
	\E[(\hat{\tau}_{\nu}(h)-\tau_\nu)^2|X_1,\dots,X_n] \approx_\P h^{2p+2-2\nu}\mathscr{B}^2 + \frac{1}{nh^{1+2\nu}} \mathscr{V},
\end{equation}
where $\approx_\P$ denotes an approximation in probability (see the SA for precise statement), and where $\mathscr{V}$ and $\mathscr{B}$ denote, respectively, approximations to the variance and bias of the $\hat{\tau}_{\nu}(h)$.

Using \eqref{eq:mse}, the MSE-optimal bandwidth choice for the RD treatment effect estimator $\hat{\tau}_{\nu}(h)$ is
\begin{equation}\label{eq:mse-h}
	h_\MSE = \left[\frac{(1+2\nu)\mathscr{V}}{2(1+p-\nu)\mathscr{B}^2}\right]^{1/(2p+3)} n^{-1/(2p+3)},
\end{equation}
where, of course, it is assumed that $\mathscr{B}\neq0$. Further details and exact formulas are given in the SA to conserve space.

The infeasible MSE-optimal bandwidth choice $h_\MSE$ can be used to construct an MSE-optimal point estimator of the RD treatment effect $\tau_\nu$, given by $\hat{\tau}_\nu(h_\MSE)$. In practice, because $\mathscr{V}$ and $\mathscr{B}$ involve unknown quantities, researchers rely on a plug-in estimator of the MSE-optimal bandwidth $h_\MSE$, say $\hat{h}_\MSE$, which is constructed by forming plug-in estimators $(\hat{\mathscr{V}}(b),\hat{\mathscr{B}}(b))$ of $(\mathscr{V},\mathscr{B})$, for some preliminary bandwidth $b\to0$; the formulas for $\hat{\mathscr{V}}(b)$ and $\hat{\mathscr{B}}(b)$ are also given in the SA. This approach gives a feasible, asymptotically MSE-optimal, RD point estimator $\hat{\tau}_\nu(\hat{h}_\MSE)$, and is commonly used in empirical work. All other MSE-optimal bandwidth choices available in the literature are also proportional to $n^{-1/(2p+3)}$, where the factor of proportionality depends on the specific MSE objective function being optimized and/or other specific methodological choices. See \citet*{Imbens-Kalyanaraman_2012_REStud}, \citet*{Calonico-Cattaneo-Titiunik_2014_ECMA}, \citet*{Arai-Ichimura_2016_EL,Arai-Ichimura_2018_QE}, and \citet*{Calonico-Cattaneo-Farrell-Titiunik_2019_RESTAT} for concrete examples, and \citet*{Cattaneo-VazquezBare_2016_ObsStud} for more general discussion.

\subsection{Robust Bias Corrected Inference}

The infeasible estimator $\hat{\tau}_\nu(h_\MSE)$ and its data-driven counterpart $\hat{\tau}_\nu(\hat{h}_\MSE)$ are MSE-optimal point estimators of $\tau_\nu$ in large samples. In empirical work, these point estimators are used not only to construct the ``best guess'' of the unknown RD treatment effect $\tau_\nu$, but also to conduct statistical inference, in particular for forming confidence intervals for $\tau_\nu$. The standard approach employs a Wald test statistic under the null hypothesis, and inverts it to form the confidence intervals. Specifically, for some choice of bandwidth $h$, the na\"ive $t$-test statistic takes the form
\[T(h) = \frac{\hat{\tau}_\nu(h) - \tau_\nu}
{\sqrt{\hat{\mathscr{V}}(h)/(n\h^{1+2\nu})}},
\]
where it is assumed that $T(h)\thicksim\mathcal{N}(0,1)$, at least in large samples, and hence the corresponding confidence interval estimator for $\tau_\nu$ is
\[I_\US(h) = \left[\;\hat{\tau}_\nu(h) - z_{1 - \frac{\alpha}{2}}\cdot \sqrt{\frac{\hat{\mathscr{V}}(h)}{n\h^{1+2\nu}}}
\;,\;
\hat{\tau}_\nu(h) - z_{\frac{\alpha}{2}}\cdot \sqrt{\frac{\hat{\mathscr{V}}(h)}{n\h^{1+2\nu}}}\;
\right], \]
where $z_\alpha$ denotes the $(100\alpha)$-percentile of the standard normal distribution. Crucially, the confidence interval $I_\US(h)$ will only have correct asymptotic coverage, in the sense of $\P[\tau_\nu \in I_\US(h)] = 1-\alpha + o(1)$, if $h$ obeys $nh^{2p+3}\to0$, that is, the bandwidth is ``small enough''. In particular, the MSE-optimal bandwidth is ``too large'':  it is easy to show that $\P[\tau_\nu \in I_\US(h_\MSE)]\not\to 1-\alpha$, rendering inference and confidence intervals based on the na\"ive $t$-test statistic $T(h_\MSE)$ invalid.

An approach to resolve the invalidity of the confidence interval $I_\US(h_\MSE)$ is to undersmooth (hence the ``$\US$'' notation) by selecting a bandwidth ``smaller'' than $h_\MSE$, or than $\hat{h}_\MSE$ in practice, when constructing the interval estimator. This approach, however, has at least two empirical and theoretical drawbacks: (i) interval length is enlarged (that is, power is decreased) because fewer observations are used, and (ii) undersmoothing is suboptimal in terms of coverage error of $I_\US(h)$. The first drawback is methodologically obvious and we will discuss it further after the new CE-optimal bandwidth choice is presented. The second drawback is formally established for RD designs in the following section as part of our main results, using novel valid coverage error expansions.

Bias correction is an alternative to undersmoothing. In the context of RD designs, \citet*{Calonico-Cattaneo-Titiunik_2014_ECMA} introduced a robust bias correction method to conduct statistical inference in general, and to form confidence intervals in particular, which in its simplest form is given as follows:
\[T_\RBC(h) = \frac{\hat{\tau}_{\nu,\BC}(h) - \tau_\nu}
{\sqrt{\hat{\mathscr{V}}_\BC(h)/(n\h^{1+2\nu})}}, \qquad
\hat{\tau}_{\nu,\BC}(h) = \hat{\tau}_\nu(h) - h^{1+p-\nu} \hat{\mathscr{B}}(b),\]
and
\[I_\RBC(h) = \left[\;\hat{\tau}_{\nu,\BC}(h) - z_{1 - \frac{\alpha}{2}}\cdot \sqrt{\frac{\hat{\mathscr{V}}_\BC(h)}{n\h^{1+2\nu}}}
\;,\;
\hat{\tau}_{\nu,\BC}(h) - z_{\frac{\alpha}{2}}\cdot \sqrt{\frac{\hat{\mathscr{V}}_\BC(h)}{n\h^{1+2\nu}}}\;
\right], \]
where again exact formulas for $\hat{\mathscr{B}}(b)$ and $\hat{\mathscr{V}}_\BC(h)$ are discussed in the SA. For inference, a key feature is that $\hat{\mathscr{V}}_\BC(h)$ is an estimator of the variance of $\hat{\tau}_{\nu,\BC}(h)$, not of the variance of $\hat{\tau}_{\nu}(h)$. For implementation, $\hat{\mathscr{B}}(b)$ depends on a local polynomial regression of order $p+1$.

An important empirical and theoretical property of $I_\RBC(h)$ is that $\P[\tau_\nu \in I_\RBC(h_\MSE)] \to 1-\alpha$, where the same bandwidth is used for both (optimal) point estimation and (suboptimal yet valid) statistical inference. Furthermore, \citet*{Calonico-Cattaneo-Titiunik_2014_ECMA} showed that the interval estimator remains valid under a wider set of bandwidth sequences, even when minimal additional smoothness of the unknown regression functions is assumed, and it was found to perform much better than other methods in both simulations and replication studies (\citealp*{Ganong-Jager_2018_JASA,Hyytinen-etal_2018_QE}). In this paper we offer principled, theoretical results that explain the good numerical properties of $I_\RBC(h)$, and we also provide new concrete ways to improve its implementation further. In the upcoming sections we present the following main results:

\begin{enumerate}[label=(\arabic*),leftmargin=*]
	
	\item We establish that $I_\RBC(h)$ has asymptotic coverage error that is no larger than  $I_\US(h)$, and is strictly smaller in most practically relevant cases, even when the corresponding best possible bandwidth is used to construct each confidence interval.
	
	\item We show that employing the MSE-optimal bandwidth $h_\MSE$ to construct $I_\RBC(h)$ is valid but suboptimal in terms of coverage error. 
	
	\item We derive new optimal bandwidth choices that minimize the coverage error of the RBC confidence intervals. We discuss the consequences for interval length and how length can be further optimized, including automatic, optimal auxiliary bandwidths. 
	
\end{enumerate}

We also discuss the implications of these results for empirical work and explore them numerically with real and simulated data.

\section{Main Results}\label{sec:mainresults}

Our main theoretical results are valid coverage error expansions for both $I_\US(h)$ and $I_\RBC(h)$. These are based on generic, valid Edgeworth expansions in the context of RD designs, which could be used for other purposes, such as studying the error in rejection probability of hypothesis tests. The generic results, and other technical details, are given in the SA. 

To state our first main result, recall that $I_\US(h)$ is constructed using $\hat{\mathscr{V}}(b)$, while $I_\RBC(h)$ is constructed using both $\hat{\mathscr{B}}(b)$ and $\hat{\mathscr{V}}_\BC(b)$, all of which are precisely described in the SA (heuristically, they are consistent estimators of higher-order biases and variances of the RD point estimator). In particular, $b$ denotes the bandwidth used to construct the bias correction estimate $\hat{\mathscr{B}}(b)$ and the associated variance estimate $\hat{\mathscr{V}}_\BC(b)$. An important quantity is $\rho=h/b$, which we discuss in detail further below.

\begin{theorem}[Coverage Error Expansions]
	\label{thm:rd}
	Suppose Assumptions \ref{ass:srd} and \ref{ass:kernel} hold, that $n \h^{1+2\nu}/ \log(n\h)^{2 + \eta} \to \infty$ for $\eta > 0$, and $\rho=h/b$ is bounded and bounded away from zero.
	\begin{enumerate}[label=(\alph*),leftmargin=*]
		
		\item If $S\geq p+1$ and $nh^{2p+3}\log(n\h)^{1 + \eta} \to 0$, then
		\begin{alignat*}{3}
			\P[\tau_\nu \in I_\US(h)] - (1 - \alpha)
			&= \frac{1}{n \h} \mathscr{Q}_{\US,1}
			&&+ n \h^{3+2p} \mathscr{Q}_{\US,2} + \h^{1+p} \mathscr{Q}_{\US,3}
			&&+ \epsilon_\US\\
			\P[\tau_\nu \in I_\RBC(h)] - (1 - \alpha)
			&= \frac{1}{n \h} \mathscr{Q}_{\RBC,1}
			&&
			&&+ \epsilon_\US,
		\end{alignat*}
		where $\epsilon_\US = o(n^{-1} \h^{-1}) + O(n \h^{3+2p+2a} + \h^{1+p+a})$.
		
		\item If $S\geq p+2$ and $nh^{2p+5}\log(n\h)^{1 + \eta} \to 0$, then
		\begin{align*}
			\P[\tau_\nu \in I_\RBC(h)] - (1 - \alpha)
			= \frac{1}{n \h} \mathscr{Q}_{\RBC,1} + n \h^{5+2p} \mathscr{Q}_{\RBC,2} + \h^{2+p} \mathscr{Q}_{\RBC,3} + \epsilon_\RBC,
		\end{align*}
		where $\epsilon_\RBC = o(n^{-1} \h^{-1}) + O(n \h^{5+2p+2a} + \h^{2+p+a})$.
	\end{enumerate}
	
	\noindent The $n$-varying, bounded quantities $(\mathscr{Q}_{\US,\ell},\mathscr{Q}_{\RBC,\ell})$, $\ell=1,2,3$, are cumbersome and hence further discussed in the SA.
	
\end{theorem}

This theorem establishes higher-order coverage error characterizations for the RD confidence intervals $I_\US(h)$ and $I_\RBC(h)$, under two distinct smoothness regimes, controlled by $S$. (Coverage error expansions under $\rho\to0$ are given in the SA because they require additional regularity conditions.) In the first case, described in part (a), the two confidence intervals are compared when the same level of smoothness is allowed. Specifically, we consider the setting where smoothness is exhausted after the leading higher-order terms of the RD point estimator $\hat{\tau}_\nu$ are characterized, which is the minimal smoothness needed to compute the MSE-optimal bandwidth $h_\MSE$, as commonly done in practice (see \eqref{eq:mse} and \eqref{eq:mse-h}). Thus, in this regime, $I_\RBC(h_\MSE)$ can be formed, but no additional smoothness is available, which gives the least favorable setting for robust bias-correction techniques. Part (a) shows, nonetheless, that even in this case, $I_\RBC(h)$ is never worse in terms of asymptotic coverage error than $I_\US(h)$, an important practical and theoretical result.

From a practical point of view, researchers first select a polynomial order (usually $p=1$), and then form confidence intervals using some bandwidth choice (often an empirical implementation of $h_\MSE$). It is rarely the case that the underlying regression functions are not smoother than what is exploited by the procedure. Part (b) discusses this case, and shows that $I_\RBC(h)$ is strictly superior to $I_\US(h)$ in terms of coverage error rates when additional smoothness is available. To be specific, comparing parts (a) and (b), it is shown that the coverage error of $I_\RBC(h)$ vanishes faster than that of $I_\US(h)$. This result gives strong theoretical justification for employing $I_\RBC(h)$ in empirical work.

The derivations in the SA also show that both $I_\US(h)$ and $I_\RBC(h)$ exhibit higher-order boundary carpentry thanks to the specific fixed-$n$ variance estimators used; see \citet*{Calonico-Cattaneo-Farrell_2018_JASA} for more discussion. This result is empirically important because it shows that the good boundary properties possessed by local polynomial estimators in point estimation carry over to inference under proper Studentization. Thus, our results formalize the crucial importance of using fixed-$n$ standard error formulas, as sometimes implemented in software for RD designs (\citealp*{Calonico-Cattaneo-Farrell-Titiunik_2017_Stata}).

Finally, the expansions given in Theorem \ref{thm:rd}, as well as the underlying technical work presented in the SA, are new to the literature. They can not be deduced from results already available (\citealp*{Calonico-Cattaneo-Farrell_2018_JASA,Calonico-Cattaneo-Farrell_2019_CEOptimal}) because they apply to the difference of two local polynomial estimates, $\hat{\tau}_{\nu}(h) = \nu!\be_\nu'\hat{\bbeta}_{+,p}(h) - \nu!\be_\nu'\hat{\bbeta}_{-,p}(h)$, and higher order terms of these differences are not trivially expressible as differences or sums of terms for each component, unlike the case when analyzing first order asymptotic approximations or MSE expansions. It is possible to upgrade the results in the SA to also show that $I_\RBC(h)$ is a coverage error optimal confidence interval estimator, uniformly over empirically-relevant classes of data generating processes, employing the optimality framework presented in \citet*{Calonico-Cattaneo-Farrell_2019_CEOptimal}. We do not provide details on this result only for brevity.

\subsection{CE-Optimal Bandwidths and Methodological Implications}

We now employ Theorem \ref{thm:rd} to develop a CE-optimal bandwidth choice for RD designs. This bandwidth choice will be made feasible in Section \ref{sec:implementation}, where we address implementation issues in detail. The following theorem is our second main result.

\begin{theorem}[Coverage Error Optimality]
	\label{thm:ce-optimal}
	Suppose the conditions of Theorem \ref{thm:rd}(b) hold. If $\mathscr{Q}_{\RBC,2}\neq0$ or $\mathscr{Q}_{\RBC,3}\neq0$, then the robust bias corrected CE-optimal confidence interval is $I_\RBC(\h_\RBC)$, where
	\[\h_\RBC = \mathscr{H} \; n^{-1/(3+p)}, \qquad
	\mathscr{H} = \argmin_{H>0} \left\vert \frac{1}{H} \mathscr{Q}_{\RBC,1}
	+ H^{5+2p} \mathscr{Q}_{\RBC,2}
	+ H^{2+p} \mathscr{Q}_{\RBC,3}\right\vert.\]
	The coverage obeys $\P[\tau_\nu \in I_\RBC(\h_\RBC)] = 1 - \alpha + O(n^{-(2+p)/(3+p)})$. 
\end{theorem}

This theorem gives the CE-optimal bandwidth choice, $\h_\RBC$, and the corresponding CE-optimal RBC confidence interval estimator, $I_\RBC(\h_\RBC)$. The optimal rate for the bandwidth sequence is $\h_\RBC\propto n^{-1/(3+p)}$, along with the associated optimal constant $\mathscr{H}$, which cannot be given in closed form (c.f. \eqref{eq:mse-h}). An analogous result is given in the SA for $I_\US(\h)$, where it is shown that the corresponding CE-optimal bandwidth choice is $\h_\US\propto n^{-1/(2+p)}$, and with a different constant of proportionality. Furthermore, this shows that $\P[\tau_\nu \in I_\US(\h_\US)] = 1 - \alpha + O(n^{-(1+p)/(2+p)})$, and therefore the RBC confidence interval estimator $I_\RBC(\h_\RBC)$ has a faster coverage error rate than the best possible undersmoothed confidence interval $I_\US(\h_\US)$. 

Our results establish that $\h_\MSE\neq\h_\RBC\neq\h_\US$ in rates (and constants, of course) for all $p\geq1$, and $\h_\MSE\asymp\h_\RBC\neq\h_\US$ for $p=0$. That is, a bandwidth different than the MSE-optimal one should be used when the goal is to construct confidence intervals with small asymptotic coverage error whenever $p\geq1$. More generally, focusing on the bandwidth choice and its consequences for coverage (interval length is addressed in the next section), we can offer three key methodological conclusions for inference in RD designs: 

\begin{enumerate}[label=(\arabic*),leftmargin=*]
	
	\item \textbf{MSE-Optimal Bandwidth}. In this case, the researcher chooses $h=\h_\MSE \propto n^{-1/(3+2p)}$. This choice of bandwidth is simple and very popular, but leads to first-order bias, rendering $I_\US(h)$ invalid. On the other hand, $T_\RBC(h_\MSE)\thicksim\mathcal{N}(0,1)$ in large samples, and hence $I_\RBC(\h_\MSE)$ is still asymptotically valid. Theorem \ref{thm:rd} quantifies the rate of coverage error decay precisely, and we find:
	\begin{align*}
		\P[\tau_\nu\in I_\US(\h_\MSE)]  - (1 - \alpha) &\asymp 1,\\
		\P[\tau_\nu\in I_\RBC(\h_\MSE)] - (1 - \alpha) &\asymp n^{-\min\{2,2+p\}/(2+p)}.
	\end{align*}
	
	\item \textbf{CE-Optimal Bandwidth for $I_\US(\h)$}. While ad-hoc undersmoothing of $\h_\MSE$ is a possible method for fixing the first-order coverage distortion of $I_\US(\h)$, a more theoretically founded choice is to use $h=\h_\US \propto n^{-1/(2+p)}$, which is also a valid choice for $I_\RBC(\h)$. In fact, this choice yields the same coverage error rate for both intervals:
	\begin{align*}
		\P[\tau_\nu\in I_\US(\h_\US)]  - (1 - \alpha) &\asymp n^{-(1+p)/(2+p)},\\
		\P[\tau_\nu\in I_\RBC(\h_\US)] - (1 - \alpha) &\asymp n^{-(1+p)/(2+p)}.
	\end{align*}
	
	\item \textbf{CE-Optimal Bandwidth for $I_\RBC(\h)$}. Finally, the researcher can also choose $h=\h_\RBC\propto n^{-1/(3+p)}$. This bandwidth choice is again too ``large'' for $I_\US(\h)$, and hence leads to a first-order coverage distortion, but is optimal for $I_\RBC(\h)$:
	\begin{align*}
		\P[\tau_\nu\in I_\US(\h_\RBC)]      - (1 - \alpha) &\asymp 1,\\
		\P[\tau_\nu\in I_\RBC(\h_\RBC)] - (1 - \alpha) &\asymp n^{-(2+p)/(3+p)}.
	\end{align*}
	
\end{enumerate}

The first point formalizes that an MSE-optimal bandwidth is always a valid choice for robust bias correction inference, with the coverage error rates depending on the polynomial order $p$. Crucially, for any $p\geq 1$, the robust bias-corrected interval $I_\RBC(\h_\MSE)$ will never achieve the fastest decay in coverage error, and therefore $\h_\MSE$ must always be undersmoothed if the goal is to constructed confidence intervals for the RD treatment effect with fastest vanishing coverage error rate. In Sections \ref{sec:implementation} and \ref{sec:extensions}, we employ this insight to propose simple rule-of-thumb CE-optimal bandwidth choices.

The last two points above reemphasize the advantages of robust bias corrected inference: $I_\US(\h_\US)$ and $I_\RBC(\h_\US)$ exhibit the same coverage error rates, which are suboptimal relative to $I_\RBC(\h_\RBC)$. In other words, $I_\RBC(\h_\RBC)$ should be preferred to all the other alternatives discussed above, when the goal is to construct CE-optimal confidence intervals in RD designs where smoothness of the underlying regression functions is not binding. This is one of the main theoretical and practical findings of this paper.

\subsection{Interval length}\label{sec:length}

An obvious concern is that the improvements in coverage offered by robust bias correction may come at the expense of larger (average) interval length. However, we now show that this is not the case. By symmetry, the (squared) length of the intervals $I_\US(h)$ and $I_\RBC(h)$ take the same form:
\[|I_\US(h)|^2 = 4\cdot z^2_{\frac{\alpha}{2}}\cdot \frac{\hat{\mathscr{V}}(h)}{n\h^{1+2\nu}}
\qquad \text{ and } \qquad
|I_\RBC(h)|^2 = 4\cdot z^2_{\frac{\alpha}{2}} \cdot \frac{\hat{\mathscr{V}}_\BC(h)}{n\h^{1+2\nu}}.\]
Thus, comparing asymptotic length amounts to examining the rate of contraction, $n^{-1}\h^{-1-2\nu}$, and the limiting variance constants, $\hat{\mathscr{V}}(h)\to_\P\mathscr{V}$ and $\hat{\mathscr{V}}_\BC(h)\to_\P\mathscr{V}_\BC$, which we show in the SA depend on the ``equivalent kernel'' function induced by the choice of $K(\cdot)$ and $\rho$ (and $p$). See \citet[Section 3.2.2]{Fan-Gijbels_1996_Book} for more discussion on equivalent kernels in local polynomial estimation.

First, regarding the contraction rate of the confidence intervals, the formal comparison follows directly from the discussion above: robust bias correction can accommodate, and will optimally employ, a slower vanishing bandwidth (i.e., $\h$ is ``larger'') than undersmoothing, and hence $I_\RBC(h)$ will contract more quickly (i.e., $n\h^{1+2\nu}\to\infty$ faster than with undersmoothing). This result formalizes the heuristic idea that using a larger bandwidth leads to more observations being used and hence improved power. To be precise, we have $|I_\RBC(\h_\RBC)|^2 \asymp n^{-(2+p)/(3+p)}$ compared to $|I_\US(\h_\US)|^2 \asymp n^{-(1+p)/(2+p)}$. It is also instructive to note that $|I_\RBC(\h_\MSE)|^2 \asymp n^{-(2p+2)/(2+p)}$ and $|\irbc(\h_\US)|^2 \asymp n^{-(1+p)/(2+p)}$, which agrees with the above discussion regarding the impact of using $\h_\MSE$, $\h_\US$, and $\h_\RBC$ to construct the interval estimators. The intervals $I_\US(\h_\MSE)$ and $I_\US(\h_\RBC)$ do not have correct asymptotic coverage.

Second, it is possible to optimize the asymptotic variance constant entering the length of the RBC confidence interval, as a function of $K(\cdot)$ and the quantity $\rho=h/b$. We can then select these two optimally to minimize the asymptotic constant portion of interval length. Specifically, \citet*{Cheng-Fan-Marron_1997_AoS} show that the asymptotic variance of a local polynomial point estimator at a boundary point is minimized by employing the uniform kernel $K(u)=\I(|u|\leq 1)$. If $I_\RBC(\h)$ is formed choosing $K(u)$ to be uniform, it follows immediately that $\rho=1$ is optimal, as with this choice the induced equivalent kernel becomes pointwise equal to the optimal equivalent kernel. For other choices of kernel $K(\cdot)$ we can derived the optimal choice of $\rho$, depending on $p$, by minimizing the $L_2$ distance between the induced equivalent kernel and the optimal variance-minimizing equivalent kernel. See the SA for all technical details.

In particular, for $\nu=0$, we computed the $L_2$-optimal $\rho$ for two popular kernels in RD applications: the triangular kernel $K(u)=(1-|u|)\I(|u|\leq 1)$, which \citet*{Cheng-Fan-Marron_1997_AoS} show is MSE-optimal (i.e., optimal from a point estimation perspective), and the Epanechnikov kernel. Table \ref{table:rho} gives the results. These $\rho^*$ optimal choices do not depend on the data, and thus are immediately implementable. For example, in the leading empirical case of $p=1$ and triangular weighting, $\rho^*=0.850$ is the recommended choice minimizing the asymptotic variance and hence the interval length of $I_\RBC(\h)$. We explore the numeric properties of these choices in Section \ref{sec:numerical}.

\section{Data-Driven Implementations}\label{sec:implementation}

We now discuss several implementable CE-optimal and related bandwidth selectors, building on our theoretical and methodological results. We focus exclusively on data-driven implementations of $I_\RBC(h)$, that is, in constructing a data-driven version of $h_\RBC$ and other related bandwidth selectors for RBC inference. We first present two main approaches to selecting the CE-optimal bandwidth choice: (i) a rule-of-thumb (ROT) based on an implementation of the MSE-optimal choice $h_\MSE$, generically denoted by $\hat{h}_\MSE$, and (ii) a direct plug-in (DPI) rule based on estimating the unknown quantities $\mathscr{Q}_{\RBC,\ell}$, $\ell=1,2,3$, and solving the optimization problem in Theorem \ref{thm:ce-optimal}. We then discuss other choices that trade-off coverage error and interval length, leveraging our coverage error expansions (Theorem \ref{thm:rd}).

The discussion below focuses on the main bandwidth $\h$, which is the crucial choice in applications. For $\rho=h/b$, i.e. the auxiliary bandwidth $b$, we consider three choices: (i) $\rho=1$, for any kernel, which corresponds to the practically relevant case $h=b$; (ii) $\rho=\rho^*$ discussed above (Table \ref{table:rho}); and (iii) $\rho=\h/\b$ estimated from the data by replacing $\h$ and $\b$ with plug-in estimators, $\hat{\h}_\MSE$ and $\hat{\b}_\MSE$, of the MSE-optimal choices for the point estimator and the bias correction, respectively. The first two choices of $\rho$ are fully automatic once $\h$ is chosen; the third requires a data-driven implementation of $\b$ as well. The form of $\b_\MSE$ can be found by selecting $(\nu,p)$ appropriately and referring to \eqref{eq:mse-h}. For example, for $\tau_0$ and $p=1$, $(\nu,p)=(2,2)$ when a quadratic approximation is used for bias correction. See the SA for details.

\subsection{ROT Bandwidth Choice}

A simple strategy to construct a feasible bandwidth selector that yields the optimal coverage error decay rate is to rescale an existing choice so that the rate agrees with $h_\RBC$. We call this the rule-of-thumb (ROT) approach. For $\hat{\h}_\MSE$ a data-driven implementation of $h_\MSE$, we simply set 
\[\hat{\h}^\mathtt{rot}_\RBC = n^{-p/((2p+3)(p+3))}\;\hat{\h}_\MSE.\]
It is immediate that $\hat{\h}^\mathtt{rot}_\RBC\propto\h_\RBC$, and therefore this empirical choice has the optimal rate of decay and yields an interval $I_\RBC(\hat{\h}^\mathtt{rot}_\RBC)$ with the fastest possible coverage error decay. As an example, for the popular local-linear RD estimator ($p=1$) and a sample of size $n=500$, the MSE-optimal bandwidth selector $\hat{\h}_\MSE$ is shrunk by $100(1-n^{-1/20})\%\approx 27\%$ to obtain RBC confidence intervals with the fastest coverage error decay rate.

Feasible MSE-optimal bandwidths are widely available in software: see \citet*{Calonico-Cattaneo-Farrell-Titiunik_2017_Stata}, and references therein, for second generation plug-in choices satisfying $\hat{\h}_\MSE/\h_\MSE\to_\P 1$. Following this, $\rho$ is selected according to the options above ($\rho=1$, $\rho=\rho^*$, or $\rho=\hat{\rho}$). It is worth noting that despite the constants being suboptimal in this approach, the ``direction'' of the trade off is still correct in the sense that if the bias is small relative to higher moments, the CE- and MSE-optimal bandwidths will increase, and $\hat{\h}^\mathtt{rot}_\RBC$ reflects this.

\subsection{DPI Bandwidth Choice}

Our second approach to constructing fully data-driven CE-optimal bandwidth choices employs plug-in estimators of the unknown constants underlying $h_\RBC$ in Theorem \ref{thm:ce-optimal}. While this bandwidth choice does not have a closed form solution in general, it is easy to form plug-in (consistent) estimators of the quantities $\mathscr{Q}_{\RBC,\ell}$, $\ell=1,2,3$, for any $\nu$, $p$, kernel, and $\rho$. Given these estimators, the DPI bandwidth selector yielding CE-optimal RBC inference is
\[\hat{\h}_\RBC = \hat{\mathscr{H}} \; n^{-1/(3+p)}, \qquad
\hat{\mathscr{H}} = \argmin_{H>0} \left\vert \frac{1}{H} \hat{\mathscr{Q}}_{\RBC,1}
+ H^{5+2p} \hat{\mathscr{Q}}_{\RBC,2}
+ H^{2+p} \hat{\mathscr{Q}}_{\RBC,3}\right\vert,\]
where $\hat{\mathscr{Q}}_{\RBC,\ell}\to_\P\mathscr{Q}_{\RBC,\ell}$, $\ell=1,2,3$, are discussed in the SA. Again, $\rho$ is chosen afterward according to the three options above ($\rho=1$, $\rho=\rho^*$, or $\rho=\hat{\rho}$). 

Estimating the quantities $\mathscr{Q}_{\RBC,\ell}$, $\ell=1,2,3$, is straightforward. These are expressed in pre-asymptotic form, so constructing the estimators boils down to replacing marginal expectations by sample averages and employing pilot bandwidth choices. Natural choices of pilot bandwidths are the corresponding MSE-optimal bandwidth selectors, already implemented in the literature. It is easy to show (see the SA for discussion) that, under regularity conditions, the DPI bandwidth selector will be consistent in the sense that $\hat{\h}_\RBC/\h_\RBC \to_\P 1$. The resulting data-driven RBC confidence intervals will be CE-optimal, given the choice of point estimator and enough smoothness of the unknown regression functions.

\subsection{Coverage Error and Interval Length Trade-Off}
\label{tradeoff}

It is natural to have a preference for shorter intervals that still have good coverage properties. Our main results allow us to discuss formally such a trade-off, and to propose alternative bandwidth choices reflecting it. Larger bandwidths (i.e., smaller values of $\gamma$ when $h=Hn^{-\gamma}$) yield on average shorter intervals: as already highlighted, one of the strengths of RBC inference is that it allows for, and will optimally employ, a larger bandwidth relative to the best undersmoothing procedure. Thus, we may seek to use a bandwidth larger than $\h_\RBC$ that reduces interval length, while still retaining good coverage properties. 

We consider the generic bandwidth choice $h_\TR = H_\TR n^{-\gamma_\TR}$, for constants $H_\TR>0$ and $\gamma_\TR>0$, where ``$\TR$'' stands for ``trade-off''. First we set the exponent $\gamma_\TR$. For valid inference, Theorem \ref{thm:rd} requires that $\gamma_\TR$ lie in $(1/(5+2p) ,  1)$ and Theorem \ref{thm:ce-optimal} gives $\h_\RBC \asymp n^{-1/(p+3)}$. For any bandwidth smaller than this ($\h \ll \h_\RBC$), both coverage error and length can be reduced with a larger bandwidth, and hence we restrict attention to:
\begin{equation}
	\label{eqn:trade_bounds}
	\frac{1}{5+2p} < \gamma_\TR \leq \frac{1}{3+p},
\end{equation}
Any choice in this range is valid in the sense that coverage error vanishes asymptotically and length is reduced compared to what $\gamma_\RBC=1/(p+3)$ would give.

To choose the constant $H_\TR$ we characterize more precisely the trade off we are making. It is perhaps not surprising that this will be about balancing, in a certain way, bias- and variance-type terms. This is also true for CE and MSE minimization, because all three methods deal with, at heart, similar fundamental quantities, but in every case the specific manifestation is different. The particulars in this case are described as follows. 

Recall from Section \ref{sec:length} that the length of $I_\RBC(h)$ does not depend on the bias, only upon the variance, and more precisely, scales as the standard deviation. Thus, squared length is proportional to variance and is therefore analogous to the first term in coverage error, which captures variance (and other centered moment) errors, but not bias. Furthermore, for the range in \eqref{eqn:trade_bounds}, the third term of coverage error is of higher order relative to the other two. Therefore, we can view a trade off of interval length against coverage error as comparing the second term of coverage error (the squared scaled bias) against a variance-type term: the square of interval length, which changes not only the constants involved but also properly adjusts for any $\nu \geq 0$ because $|I_\RBC(h)|^2 \asymp n^{-1}h^{-1-2\nu}$. The leading constant portions of coverage error and length are $H^{5+2p} \hat{\mathscr{Q}}_{\RBC,2}$ and $4 z^2_{\frac{\alpha}{2}} \hat{\mathscr{V}}_\BC H^{-1-2\nu}$, respectively, where $\hat{\mathscr{V}}_\BC$ and $\hat{\mathscr{Q}}_{\RBC,2}$ are preliminary feasible estimators of $\mathscr{V}_\BC$ and $\mathscr{Q}_{\RBC,2}$. Therefore, we select the constant $H_\TR$ in $h_\TR = H_\TR n^{-\gamma_\TR}$ as
\begin{align*}
	\hat{H}_\TR & = \argmin_{ H > 0} \ W \times H^{5+2p} \hat{\mathscr{Q}}_{\RBC,2} + (1-W) \times H^{-1-2\nu} 4 z^2_{\frac{\alpha}{2}} \hat{\mathscr{V}}_\BC    		 \\
	& = \left(\frac{1-W}{W}\frac{1+2\nu}{5+2p} \frac{4 z_{\frac{\alpha}{2}}^2 \hat{\mathscr{V}}_\BC }{\hat{\mathscr{Q}}_{\RBC,2}}\right)^{\frac{1}{6 + 2p+2\nu}},			
\end{align*}
for a researcher-chosen weight $W \in (0,1)$. In Section \ref{sec:numerical} we find that $\hat{h}_\TR = \hat{H}_\TR n^{-\gamma_\TR}$ (and its infeasible counterpart $h_\TR$) behaves as expected, with the natural choice of $W=1/2$ and $\gamma_\TR=0.1964$, the midpoint of \eqref{eqn:trade_bounds} for $p=1$.

\section{Extensions}\label{sec:extensions}

We briefly discuss several extensions of our main results. Unlike results based on first-order asymptotic approximations, establishing valid higher-order Edgeworth expansions in the settings of this section would require non-trivial additional work beyond the scope of this paper. Nevertheless, following the logic and results above, we can provide simple ROT bandwidth choices targeting inference, based on already-available MSE-optimal bandwidth selectors.

\subsection{Other RD Designs}

In the context of fuzzy (and fuzzy kink) RD designs, the estimand and estimator are ratios of sharp RD design estimands and estimators, respectively.  First-order asymptotic approximations follow directly from standard linearization methods, and although the validity of the coverage error expansion can be similarly proven to hold, this is no help in computing the terms of the expansion. That is, even though the linearization error has no effect on the first-order asymptotic approximation, it can have a direct effect on the Edgeworth and coverage error expansions. Without capturing the effect of the linearization, full derivation of inference optimal bandwidths is not possible. However, in this context we propose the following ROT bandwidth:
\[\breve{\h}^\mathtt{rot}_\RBC = n^{-p/((2p+3)(p+3))}\;\breve{\h}_\MSE,\]
where $\breve{\h}_\MSE$ denotes an implementation of the MSE-optimal bandwidth for the fuzzy (or fuzzy kink) RD estimator. Sharp, fuzzy, and kink RD designs also arise in geographic, multi-score, and multi-cutoff RD settings (\citealp*{Papay-Willett-Murnane_2011_JoE,Keele-Titiunik_2015_PA,Cattaneo-Keele-Titiunik-VazquezBare_2016_JOP}), and the results in this paper can also be used in those cases directly.

\subsection{Clustered Data}

When the data exhibits clustering, first-order asymptotic results can be easily extended to account for clustered sampling where (i) each unit $i$ belongs to exactly one of $G$ clusters and (ii) $G \to \infty$ and $G h \to \infty$ (see \citet*{Bartalotti-Brummet_2017_AIE} and \citet*{Calonico-Cattaneo-Farrell-Titiunik_2019_RESTAT}). Since MSE-optimal bandwidth choices in this context are available and fully implemented, the corresponding ROT implementation is:
\[\check{\h}^\mathtt{rot}_\RBC = G^{-p/((2p+3)(p+3))}\;\check{\h}_\MSE,\]
where now $G$ denotes the number of clusters, and $\check{\h}_\MSE$ denotes an implementation of the MSE-optimal bandwidth accounting for clustering. Robust bias-corrected confidence intervals ar formed using this bandwidth choice, together with appropriate (cluster-robust) standard error estimators.

\subsection{Pre-intervention Covariates}
\label{covs}

\citet*{Calonico-Cattaneo-Farrell-Titiunik_2019_RESTAT} employs first-order asymptotics to characterize formally the implications of including pre-intervention covariates in the estimation of and inference for RD treatment effects. Again, this is not sufficient for higher order expansions and the inclusion of covariates will impact the coverage error expansion, making it impossible to derive a fully optimal bandwidth from existing results. However, a ROT bandwidth selector in this context is:
\[\tilde{\h}^\mathtt{rot}_\RBC = n^{-p/((2p+3)(p+3))}\;\tilde{\h}_\MSE,\]
where $n$ denotes the sample size and $\tilde{\h}_\MSE$ denotes an implementation of the MSE-optimal bandwidth accounting for the inclusion of additional pre-intervention covariates. Robust bias-corrected confidence intervals are formed using this bandwidth choice, together with appropriate covariate-adjusted standard error estimators.

\section{Numerical Results}\label{sec:numerical}

We present empirical evidence highlighting the performance of the new RD bandwidth selection and inference methods developed. We consider a Monte Carlo experiment and an empirical application, both employing the dataset of \citet*{Ludwig-Miller_2007_QJE} used to study the effect of Head Start assistance on child mortality. This canonical dataset was employed before by \citet*{Calonico-Cattaneo-Titiunik_2014_ECMA}, \citet*{Cattaneo-Titiunik-VazquezBare_2017_JPAM} and \citet*{Calonico-Cattaneo-Farrell-Titiunik_2019_RESTAT}, where further institutional and descriptive information is provided.

\subsection{Monte Carlo Experiment}

The simulations use $n=500$ i.i.d.\ draws, $i=1,2,...,n$, from the model
\begin{equation*}
	Y_i=m(X_i)+\varepsilon_i,\qquad X_i\thicksim 2\mathcal{B}(2,4)-1,\qquad \varepsilon_i\thicksim \mathcal{N}(0,\sigma_{\varepsilon}^2)
\end{equation*}
where $\mathcal{B}(\alpha ,\beta )$ denotes a beta distribution with parameters $\alpha $ and $\beta $, and the regression function $m(x)$ is obtained from the Head Start data. Specifically, we estimate the regression function using a $5$-th order polynomial with separate coefficients for $X_i<\x$ and $X_i>\x$, where $X_i$ is a poverty index and $\x=59.1984$ is the RD cutoff point. This estimation leads to:
\begin{equation*}
	m(x) =\begin{cases}
		3.71 + 2.30x  + 3.28 x^2 + 1.45 x^3 + 0.23 x^4 + 0.03x^5 & \text{if } x<\x \\
		0.26 + 18.49x - 54.81x^2 + 74.30x^3 - 45.02x^4 + 9.83x^5 & \text{if } x\geq\x
	\end{cases},
\end{equation*}
with $\sigma_{\varepsilon}=0.6136$.

We consider $5,000$ replications, and report empirical coverage and average interval length for a variety of inference procedures. Specifically, Table \ref{table:simuls} considers undersmoothing ($I_\US(\h)$) and robust bias-corrected ($I_\RBC(\h)$) confidence intervals for different choices of bandwidths $\h$ and parameter $\rho$. In all cases we consider a local-linear RD estimator ($p=1$) with the triangular kernel and ``HC3'' heteroskedasticity consistent standard errors, motivated by the fact that the least-squares residuals are on average too small (see the SA for more). 

The results in Table \ref{table:simuls} are organized as follows. The table presents three groups by row: (i) procedures employing MSE-optimal bandwidth choices ($h_\MSE,\hat{h}_\MSE,\tilde{h}_{\MSE}$), (ii) procedures employing CE-optimal bandwidth choices ($h_\RBC,\hat{h}_\RBC,h^\texttt{rot}_\RBC,\hat{h}^\texttt{rot}_\RBC,\tilde{h}^\texttt{rot}_{\RBC}$), and (iii) procedures employing trade-off bandwidth choices ($h_\TR,\hat{h}_\TR$). Quantities without hats or tildes correspond to infeasible bandwidth choices, quantities with hats denote feasible implementations (DPI without label, or ROT with corresponding label), and quantities with tilde denote feasible implementations with covariate adjustment. For the latter the model includes, as a predetermined covariate, percentage of urban population in 1960. 

The table also presents three groups by columns: (i) ``Bandwidth'' reports infeasible or average feasible bandwidth choices (recall $\hat{\rho}=\hat{\h}/\hat{\b}_\MSE$ with $\hat{\h}$ as appropriate); (ii) ``Empirical Coverage'' reports coverage of $I_\US(\h)$ and of $I_\RBC(\h)$ for three choices of $\rho=\h/\b$; and (iii) ``Interval Length'' reports the average length of the same four distinct confidence intervals (undersmoothing and three implementations of RBC indexed by the choice of $\rho$). Further implementation details are given in the SA. 

All the findings emerging from the simulation study are in qualitative agreement with the main theoretical results from our paper. Confidence intervals based on undersmoothing, $I_\US(\h)$, did not exhibit good coverage properties, while those based on RBC, $I_\RBC(\h)$, performed well. While the MSE-optimal bandwidth selectors also worked well, the CE-optimal bandwidth selectors offered some empirical refinements in terms of coverage error. Furthermore, the bandwidth selector based on coverage error and interval length trade-off discussed in Section \ref{tradeoff} also performed well. Other empirical findings are in line with our theoretical and methodological discussions.

\subsection{Empirical Application}

To complement the Monte Carlo experiment, we also employed the Head Start data to illustrate the performance of our new bandwidth selection and inference methods using a realistic empirical application. Specifically, we study the RD treatment effect of Head Start assistance on child mortality following the original work of \citet*{Ludwig-Miller_2007_QJE}. See also \citet*{Cattaneo-Titiunik-VazquezBare_2017_JPAM} for a recent re-examination of the empirical findings using modern RD methodology.

In this application, the unit of analysis is a U.S. county, and eligibility into Head Start assistance was based on each county's poverty index in $1960$. The RD design naturally emerges by the assignment rule to the program: $T_i=\I(X_i\geq \x)$, where $X_i$ denotes the $1960$ poverty index of county $i$ and $\x=59.1984$ was the federally-mandated cutoff point. The outcome variable considered is mortality rates per $100,000$ for children between $5$--$9$ years old, with Head Start-related causes, during the period $1973$--$1983$.

The main empirical results are presented in Table \ref{table:empapp}. We first report the sharp RD treatment effect estimator using a local-linear estimator ($p=1$) with triangular kernel and MSE-optimal bandwidth. In line with previous findings, we obtain $\hat{\tau}_0(\hat{\h}_\MSE)=-2.409$ with $\hat{\h}_\MSE=6.81$. Next, we compute several RBC confidence intervals with different choices of bandwidths $h$ and $\rho$, including the new inference procedures proposed in this paper. In all cases, the empirical results are in qualitatively agreement and confirm an RD treatment effect that is statistically different from zero.

\section{Conclusion}\label{sec:conclusion}

This paper presented two main results for RD designs, which have concrete practical implications for empirical work. First, we established valid coverage error expansions of na\"ive and robust bias-corrected confidence intervals for RD treatment effects, and showed that the latter confidence intervals never have asymptotically larger coverage errors and can indeed offer higher-order refinements whenever the underlying regression functions are smooth enough (arguably the most relevant case in applications). Thus, this result offers concrete guidance for empirical work in RD designs by ranking competing confidence interval estimators encountered in practice.

Second, using our coverage error expansions, we also developed CE-optimal bandwidth choices and discussed how to implement them in practice. The same way that MSE-optimal bandwidths deliver MSE-optimal point estimators for RD treatment effects, our new CE-optimal bandwidth choices deliver inference-optimal  confidence intervals in the sense that their coverage error is the smallest possible given the choice of point estimator used. This second result also offers concrete empirical guidance for applied work using RD designs, providing a companion bandwidth choice to be used when forming confidence intervals for RD treatment effects.

\bibliography{Calonico-Cattaneo-Farrell_2020_ECTJ--Bibliography}
\bibliographystyle{chicago}

\clearpage

\begin{table}[h]
	{\begin{center}
			\caption{$L_2$-Optimal $\rho$}\label{table:rho}
			{
\begin{tabular}{ccccc}
\hline\hline
\multicolumn{1}{c}{\bfseries $p$}&\multicolumn{1}{c}{\bfseries }&\multicolumn{3}{c}{\bfseries Kernel}\tabularnewline
\cline{1-1} \cline{3-5}
\multicolumn{1}{c}{}&\multicolumn{1}{c}{}&\multicolumn{1}{c}{Triangular}&\multicolumn{1}{c}{Epanechnikov}&\multicolumn{1}{c}{Uniform}\tabularnewline
\hline
$0$&&$0.778$&$0.846$&$1$\tabularnewline
$1$&&$0.850$&$0.898$&$1$\tabularnewline
$2$&&$0.887$&$0.924$&$1$\tabularnewline
$3$&&$0.909$&$0.940$&$1$\tabularnewline
$4$&&$0.924$&$0.950$&$1$\tabularnewline
\hline
\end{tabular}
}\bigskip
	\end{center}}
	\vspace{-.2in}\footnotesize\textbf{Note}: Computed by minimizing the $L_2$ distance between the RBC induced equivalent kernel and the optimal variance-minimizing equivalent kernel obtained by \citet*{Cheng-Fan-Marron_1997_AoS} for $\nu=0$.
\end{table}

\begin{table}[h]
	{\begin{center}
			\caption{Empirical Coverage and Average Interval Length of 95$\%$ Confidence Intervals}\label{table:simuls}
			\resizebox{\textwidth}{!}{
\begin{tabular}{lcccccccccccccc}
\hline\hline
\multicolumn{1}{l}{\bfseries }&\multicolumn{2}{c}{\bfseries Bandwidth}&\multicolumn{1}{c}{\bfseries }&\multicolumn{5}{c}{\bfseries Empirical Coverage}&\multicolumn{1}{c}{\bfseries }&\multicolumn{5}{c}{\bfseries Interval Length}\tabularnewline
\cline{2-3} \cline{5-9} \cline{11-15}
\multicolumn{1}{l}{}&\multicolumn{1}{c}{$h$}&\multicolumn{1}{c}{$\widehat{\rho}$}&\multicolumn{1}{c}{}&\multicolumn{1}{c}{US}&\multicolumn{1}{c}{RBC:}&\multicolumn{1}{c}{$\widehat{\rho}$}&\multicolumn{1}{c}{$\rho^*$}&\multicolumn{1}{c}{$\rho=1$}&\multicolumn{1}{c}{}&\multicolumn{1}{c}{US}&\multicolumn{1}{c}{RBC:}&\multicolumn{1}{c}{$\widehat{\rho}$}&\multicolumn{1}{c}{$\rho^*$}&\multicolumn{1}{c}{$\rho=1$}\tabularnewline
\hline
&&&&&&&&&&&&&&\tabularnewline
$h_{\tt MSE}$&0.154&0.520&&92.7&&94.5&93.6&92.8&&1.14&&1.28&1.50&1.64\tabularnewline
$\widehat{h}_{\tt MSE}$&0.174&0.571&&88.7&&93.7&93.4&93.0&&1.08&&1.24&1.42&1.55\tabularnewline
$\widetilde{h}_{\tt MSE}$&0.173&0.571&&88.9&&93.7&93.6&93.0&&1.08&&1.24&1.43&1.56\tabularnewline
\hline
&&&&&&&&&&&&&&\tabularnewline
$h_{\tt RBC}$&0.145&0.492&&93.1&&94.6&93.5&92.6&&1.17&&1.31&1.55&1.69\tabularnewline
$\widehat{h}_{\tt RBC}$&0.163&0.535&&88.7&&93.6&91.0&90.5&&1.15&&1.32&1.50&1.64\tabularnewline
$h^{\tt rot}_{\tt RBC}$&0.113&0.381&&94.0&&94.7&93.0&91.9&&1.35&&1.46&1.79&1.94\tabularnewline
$\widehat{h}^{\tt rot}_{\tt RBC}$&0.127&0.418&&92.4&&94.3&93.4&92.2&&1.28&&1.39&1.69&1.83\tabularnewline
$\widetilde{h}^{\tt rot}_{\tt RBC}$&0.127&0.416&&92.5&&94.3&93.1&92.1&&1.28&&1.39&1.69&1.84\tabularnewline
\hline
&&&&&&&&&&&&&&\tabularnewline
$h_{\tt TO}$&0.203&0.686&&88.8&&94.2&93.9&93.4&&0.98&&1.19&1.30&1.42\tabularnewline
$\widehat{h}_{\tt TO}$&0.172&0.566&&87.3&&93.6&90.9&90.7&&1.11&&1.31&1.46&1.59\tabularnewline
\hline
\end{tabular}
}\bigskip
	\end{center}}
	\vspace{-.2in}\footnotesize\textbf{Note}: US denotes undersmoothed confidence interval, $I_\US(h)$, and RBC denotes robust bias-corrected confidence interval, $I_\RBC(h)$. Procedures are computed using the triangular kernel, $p=1$, and HC$_3$ variance estimation. Recall that $\hat{\rho}=\hat{\h}/\hat{\b}_\MSE$ for corresponding bandwidth selectors $\hat{\h}$ (given in the table), and $\rho^*=0.850$ (Table \ref{table:rho}).
\end{table}

\begin{table}[h]
	{\begin{center}
			\caption{Head Start Empirical Application}\label{table:empapp}
			\resizebox{\textwidth}{!}{
\begin{tabular}{lcccccccc}
\hline\hline
\multicolumn{1}{l}{\bfseries }&\multicolumn{1}{c}{\bfseries Point Estimate}&\multicolumn{1}{c}{\bfseries }&\multicolumn{2}{c}{\bfseries Bandwidth}&\multicolumn{1}{c}{\bfseries }&\multicolumn{3}{c}{\bfseries RBC Confidence Intervals}\tabularnewline
\cline{2-2} \cline{4-5} \cline{7-9}
\multicolumn{1}{l}{}&\multicolumn{1}{c}{}&\multicolumn{1}{c}{}&\multicolumn{1}{c}{$\widehat{h}$}&\multicolumn{1}{c}{$\widehat{\rho}$}&\multicolumn{1}{c}{}&\multicolumn{1}{c}{$\widehat{\rho}$}&\multicolumn{1}{c}{$\rho^*$}&\multicolumn{1}{c}{$\rho=1$}\tabularnewline
\hline
&&&&&&&&\tabularnewline
$\widehat{h}_{\tt MSE}$&-2.409&&6.81&0.635&&[-5.46 , -0.1]&[-5.88 , -0.44]&[-6.41 , -1.09]\tabularnewline
$\widetilde{h}_{\tt MSE}$&-2.473&&6.98&0.651&&[-5.21 , -0.37]&[-5.78 , -0.69]&[-6.54 , -1.39]\tabularnewline
\hline
&&&&&&&&\tabularnewline
$\widehat{h}_{\tt RBC}$&-3.311&&4.467&0.416&&[-6.14 , -0.82]&[-6.53 , -1.09]&[-6.23 , -0.27]\tabularnewline
$\widehat{h}^{\tt rot}_{\tt RBC}$&-3.273&&4.581&0.427&&[-6.12 , -0.78]&[-6.57 , -1.16]&[-6.26 , -0.39]\tabularnewline
$\widetilde{h}^{\tt rot}_{\tt RBC}$&-3.526&&4.696&0.438&&[-6.13 , -1.25]&[-7.06 , -1.75]&[-6.93 , -1.23]\tabularnewline
\hline
\end{tabular}
}\bigskip
	\end{center}}
	\vspace{-.2in}\footnotesize\textbf{Note}: Procedures are computed using the triangular kernel, $p=1$, and HC$_3$ variance estimation. Recall that $\hat{\rho}=\hat{\h}/\hat{\b}_\MSE$ for bandwidth selectors $\hat{\h}$ (given in the table), and $\rho^*=0.850$ (Table \ref{table:rho}).
\end{table}

\end{document}


\begin{abstract}

        \noindent This supplement contains technical details and formulas omitted from the main text, proofs of all theoretical results, further technical and methodological derivations, and details on practical and numerical implementations.

    \end{abstract}

\section{Setup, Notation, and Assumptions}
\label{supp:setup}

We assume the researcher observes a random sample $(Y_i,T_i,X_i)'$, $i=1,2,\dots,n$, where $Y_i$ denotes the outcome variable of interest, $T_i$ denotes treatment status, and $X_i$ denotes an observed continuous score or running random variable, which determines treatment assignment for each unit in the sample. In the canonical sharp RD design, all units with $X_i$ not smaller than a known threshold $\x$ are assigned to the treatment group and take-up treatment, while all units with $X_i$ smaller than $\x$ are assigned to the control group and do not take-up treatment, so that $T_i=\I(X_i\geq\x)$. Using the potential outcomes framework, $Y_i=Y_i(0)\cdot(1-T_i)+Y_i(1)\cdot T_i$, with $Y_i(1)$ and $Y_i(0)$ denoting the potential outcomes with and without treatment, respectively, for each unit.

The parameter of interest in sharp RD designs are either the average treatment effect at the cutoff or its derivatives. Thus, herein we study the generic population parameter, for some integer $\v \geq 0$:
\begin{equation}
\label{suppeqn:tau}
\tau_\nu = \tau_\nu(\x) 
= \left.\frac{\partial^\nu}{\partial x^\nu}\E[Y_i(1)-Y_i(0)|X_i=x]\right|_{x=\x},
\end{equation}

Here and elsewhere we drop evaluation points of functions when it causes no confusion. With this notation, $\tau_0$ corresponds to the standard sharp RD estimand, while $\tau_1$ denotes the sharp kink RD estimand (up to scale).

\subsection{Local Polynomial Point Estimation}
\label{supp:estimators}

Will not give a complete treatment of local polynomial estimation here. For background, careful derivations of the results and formulas herein, and further technical details, see the following: \citet*{Fan-Gijbels_1996_Book} for background, \citet*{Calonico-Cattaneo-Titiunik_2014_ECMA} in the context of RD specifically, and \citet*{Calonico-Cattaneo-Farrell_2018_JASA,Calonico-Cattaneo-Farrell_2019_CEOptimal} for further technical details particularly in the context of Edgeworth expansions.

We estimate $\tau_\nu$ by taking the difference of two local polynomial estimators, from each side of $\x$. Define the coefficients of the (one-sided, weighted) local regressions:
\begin{align}
\begin{split}
\label{suppeqn:beta hat}
\bhat_- = \bhat_{-,p}(h) & = \argmin_{\bb \in \mathbb{R}^{p+1}} \sumi  ( Y_i - \br_p(X_i - \x)'\bb)^2  K_- \left( \Xhi \right)  = \frac{1}{n \h^\v} \Gp{-}^{-1} \Op{-}  \bY,     		\\
\bhat_+ = \bhat_{-,p}(h) & = \argmin_{\bb \in \mathbb{R}^{p+1}} \sumi   ( Y_i - \br_p(X_i - \x)'\bb)^2  K_+ \left( \Xhi \right)  = \frac{1}{n \h^\v} \Gp{+}^{-1} \Op{+}  \bY,  
\end{split}
\end{align}
where:
\begin{itemize}
	
	\item $p$ is an integer greater than $\min\{1,\v\}$,
	
	\item $\be_k$ is a conformable zero vector with a one in the $(k+1)$ position, for example $\be_\v$ is the $(p+1)$-vector with a one in the $\v^{\text{th}}$ position and zeros in the rest, 
	
	\item $\br_p(u) = (1, u, u^2, \ldots, u^p)'$,
	
	\item $\h$ is a positive bandwidth sequence that vanishes as $n$ diverges,
	
	\item $\Xhi = (X_i - \x)/\h$, for a bandwidth $\h$ and point of interest $\x$,
	
	\item $K_-(u) = \One\{u < 0\}K(u)$ and $K_+(u) = \One\{u \geq 0\}K(u)$ for a kernel function $K(u)$, with in particular $ K_- ( \Xhi ) = \One(X_i < \x) K ( \Xhi )$ and $K_+ ( \Xhi ) = \One(\x \leq X_i) K ( \Xhi )$, 
	
	\item to save space, products of functions will often be written together, with only one argument, for example
	\[ (K \br_p \br_p')(\Xhi) \defsym K(\Xhi) r_p(\Xhi) r_p(\Xhi)' = K\left(\frac{X_i - \x}{\h}\right) \br_p \left(\frac{X_i - \x}{\h}\right) \br_p \left(\frac{X_i - \x}{\h}\right)' ,  \]
	
	\item $\Gp{-} = \frac{1}{n\h} \sumi (K_- \br_p \br_p')(\Xhi)$ and $\Gp{+} = \frac{1}{n\h} \sumi (K_+ \br_p \br_p')(\Xhi)$,
	
	\item $\Op{-} = \h^{-1} [ (K_- \br_p)(X_{\h,1}), \ldots,  (K_- \br_p)(X_{\h,n})]$ and $\Op{+} = \h^{-1} [ (K_+ \br_p)(X_{\h,1}),  \ldots,  (K_+ \br_p)(X_{\h,n})]$, and
	
	\item $\bY = (Y_1, \ldots, Y_n)'$.
	
\end{itemize}

We maintain the same bandwidth and kernel function on both sides of the cutoff for notational simplicity. Accommodating different bandwidths, which share a rate of decay, is only a matter of notational burden. At the expense of substantial complication, any aspect of the local polynomial fit on one side can be different from the other, including the bandwidth rate and the order $p$; all the results will still hold in principle. As this approach is rarely taken in practice, we decide not to introduce the complication.

The standard point estimator of the parameter of interest $\tau_\v$ of Equation \eqref{suppeqn:tau} is then the difference of the appropriate two entries from the one-sided coefficient vectors:
\begin{equation}
\label{suppeqn:tau hat}
\that_\nu = \that_{\v,\US} =  \hat{\tau}_{\nu}(h) = \v!\be_\nu'\hat{\bbeta}_+ - \v!\be_\nu'\hat{\bbeta}_- =  \frac{1}{n\h^\v} \v!\be_\nu'\left( \Gp{+}^{-1} \Op{+}  -  \Gp{-}^{-1} \Op{-} \right) \bY,
\end{equation}
which is also denoted $\that_{\v,\US}$ to explicitly refer to the fact that undersmoothing is required for valid inference. Compared to the main text, we will often drop the dependence on the bandwidth unless it is required to make a specific point.

\subsection{Assumptions}
\label{supp:asmpts}

Let $g^{(s)}(x)=\partial^\nu g(x)/\partial x^\nu$ for any sufficiently smooth function $g(\cdot)$, with $g(x)=g^{(0)}(x)$ to save notation.

\begin{assumption}[RD]\label{suppasmpt:srd}
	For some $S > p \geq \min\{1,\v\}$ and all $x\in[x_l,x_u]$, where $x_l<\x<x_u$,
	\begin{enumerate}
		\item the Lebesgue density of $X_i$, denoted by $f(x)$, is positive and continuous,
		\item $\mu_-(x) = \E[Y_i(0)|X_i=x]$ and $\mu_+(x) = \E[Y_i(1)|X_i=x]$ are $\S$ times continuously differentiable, with $\mu_-^{(\S)}(x)$ and $\mu_+^{(\S)}(x)$ H\"older continuous with exponent $\s\in(0,1]$, and
		\item $\E[|Y_i(t)|^\delta|X_i=x]$ continuous, for $t\in\{0,1\}$ and $\delta>8$, with $\sigma^2_-(x) = \V[Y_i(0)|X_i=x]$ and $\sigma^2_+(x) = \V[Y_i(1)|X_i=x]$ positive and continuous, and
		\item the Lebesgue density of $(Y(t), X)$, $f_{y_t x}(\cdot)$, is positive and continuous.
	\end{enumerate}
\end{assumption}

The only difference between this assumption and its counterpart in the main text is that we have defined the function $\mu_+$, $\mu_-$, $\sigma^2_+(x)$, and $\sigma^2_-(x)$, which we will need later. With this notation the parameter of interest is (cf. \eqref{suppeqn:tau})
\begin{equation*}
\tau_\nu = \tau_\nu(\x) 
= \left.\frac{\partial^\nu}{\partial x^\nu}\E[Y_i(1)-Y_i(0)|X_i=x]\right|_{x=\x}          = \mu^{(\nu)}_{+}(\x)-\mu^{(\nu)}_{-}(\x)
\end{equation*}
and the standard point estimator is (cf. \eqref{suppeqn:tau hat})
\begin{equation*}
\that_\nu = \that_{\v,\US} =  \hat{\tau}_{\nu}(h) = \v!\be_\nu'\hat{\bbeta}_+ - \v!\be_\nu'\hat{\bbeta}_- = \hat{\mu}^{(\nu)}_{+}(\x) - \hat{\mu}^{(\nu)}_{-}(\x).
\end{equation*}

The conditions on the kernel function are as follows.

\begin{assumption}[Kernel]\label{suppasmpt:kernel}
	$K(u)=\I(u<0)k(-u)+\I(u\geq 0)k(u)$, where $k(\cdot):[0,1]\mapsto\mathbb{R}$ is bounded and continuous on its support, positive $(0,1)$, zero outside its support, and either is constant or $(1, K(u) \br_{3(p+1)}(u)')$ is linearly independent on $(0,1)$.
\end{assumption}

\section{Technical Details and Formulas Omitted from the Main Text}
\label{supp:details}

In this section we state formulas and technical details omitted from the main text. These consist of bias and variance terms and their estimators and the terms of the coverage error expansion. Throughout we maintain Assumptions \ref{suppasmpt:srd} and \ref{suppasmpt:kernel} with $\S \geq p+1$, or, where mentioned, $\S \geq p + 2$. Derivations of many of the formulas in the first two subsections can be found in \citet*{Calonico-Cattaneo-Titiunik_2014_ECMA}. When sufficient smoothness does not exist, the results of \citet*{Calonico-Cattaneo-Farrell_2018_JASA,Calonico-Cattaneo-Farrell_2019_CEOptimal} apply.

Recall from Equation \eqref{suppeqn:tau hat} that the standard point estimator of the parameter of interest $\tau_\v$ of Equation \eqref{suppeqn:tau} is the difference of the appropriate two entries from the one-sided coefficient vectors, 
\begin{equation*}
\that_\nu = \hat{\tau}_{\nu}(h) = \v!\be_\nu'\hat{\bbeta}_+ - \v!\be_\nu'\hat{\bbeta}_- =  \frac{1}{n\h^\v} \v!\be_\nu'\left( \Gp{+}^{-1} \Op{+}  -  \Gp{-}^{-1} \Op{-} \right) \bY,
\end{equation*}
which will also be denoted $\that_{\v,\US}$ to explicitly refer to the fact that undersmoothing is required for valid inference.

\subsection{Bias and Bias Correction}
\label{supp:bias}

The conditional bias of $\that_\v$ obeys
\begin{align}
\begin{split}
\label{suppeqn:bias us}
& \E\left[\that_\nu \big| X_1, \ldots, X_n \right] - \tau_\v  =   \h^{p+1 - \v}  \mathscr{B}   +   o_P( \h^{p+1 - \v}),  		\\
& \text{where} \qquad 	\mathscr{B}  = \frac{\v!}{(p+1)!} \be_\v' \left( \Gp{+}^{-1} \Lp{+} \mu_+^{(p+1)} - \Gp{-}^{-1} \Lp{-} \mu_-^{(p+1)}\right),
\end{split}
\end{align}
with
\begin{itemize}
	\item $\Lp{+} = \Op{+}\left[ X_{\h,1}^{p+1}, \ldots,  X_{\h,n}^{p+1}\right]'/n$ and similarly for $\Lp{-}$, and 
	\item $\mu_+^{(p+1)} = \left.\frac{\partial^\v}{\partial x^\v} \E[Y(1) \vert X_i = x] \right|_{x = \x}$, and similarly for $\mu_-^{(p+1)}$, see Assumption \ref{suppasmpt:srd}.
\end{itemize}

The bias of \eqref{suppeqn:bias us} is first-order important without further steps. See the main paper for discussion. Because its asymptotic order is $\h^{p+1 - \v}$, undersmoothing relies on a ``small'' bandwidth choice, i.e. one assumed to vanish rapidly enough to render this bias ignorable. Robust bias correction involves estimating $\mathscr{B}$ and subtracting this estimate from the point estimator $\that_\v$. The estimator of $\mathscr{B}$ will also be based on one-sided local polynomial regressions, of exactly the same form as \eqref{suppeqn:beta hat} but with the degree of the polynomial one order higher, $q=p+1$ (see Remark \ref{supprem:higher q}), and a bandwidth $\b$ defined as $\b = \rho^{-1}\h$. Specifically,
\begin{equation}
\label{suppeqn:that rbc}
\that_{\v,\BC} = \that_\v - \h^{p+1 - \v}  \hat{\mathscr{B}} =  \frac{1}{n\h^\v} \v!\be_\nu'\left( \Gp{+}^{-1} \Orbc{+}  -  \Gp{-}^{-1} \Orbc{-} \right) \bY,
\end{equation}
where
\begin{equation}
\label{suppeqn:bias hat}
\hat{\mathscr{B}} = \frac{\v!}{(p+1)!} \be_\v' \left( \Gp{+}^{-1} \Lp{+} \hat{\mu}_+^{(p+1)} - \Gp{-}^{-1} \Lp{-} \hat{\mu}_-^{(p+1)}\right),
\end{equation}
and
\[
\Orbc{+} = \Op{+} - \rho^{p+1} \Lp{+} \be_{p+1}' \Gq{+}^{-1} \Oq{+}  		  \quad \text{ and } \quad  		  \Orbc{-} = \Op{-} - \rho^{p+1} \Lp{-} \be_{p+1}' \Gq{-}^{-1} \Oq{-}
\]
stemming from the estimation of the derivatives using local polynomials. That is,
\[ \hat{\mu}_+^{(p+1)}  =  \frac{1}{n\b^{p+1}} (p+1)!\be_{p+1}' \Gq{+}^{-1} \Oq{+} \bY  		  \quad \text{ and } \quad  		  \hat{\mu}_-^{(p+1)}  =  \frac{1}{n\b^{p+1}} (p+1)!\be_{p+1}' \Gq{-}^{-1} \Oq{-} \bY,    \]
with
\begin{itemize}
	
	\item an integer $q \geq p$ taken throughout to be $q = p+1$ (\citet*{Calonico-Cattaneo-Farrell_2018_JASA} show why $q=p+1$ is the optimal choice for coverage considerations. See also Remark \ref{supprem:higher q}) and 
	
	\item $\b = \rho^{-1} \h$ is a positive bandwidth sequence that vanishes as $n$ diverges.
\end{itemize}
Given these, the rest of the notation is defined analogously to the above, namely:
\begin{itemize}
	
	\item $\br_q(u) = (1, u, u^2, \ldots, u^q)'$,	
	
	\item $\Xbi = (X_i - \x)/\b$, for a bandwidth $\b$ and point of interest $\x$,
	
	\item $\Gq{-} = \frac{1}{n\b} \sumi (K_- \br_q \br_q')(\Xbi)$ and $\Oq{-} = \b^{-1} [ (K_- \br_q)(X_{\b,1}), \ldots,  (K_- \br_q)(X_{\b,n})]$ and similarly for $\Gq{+}$ and $\Oq{+}$. 	
	
\end{itemize}

The bias of $\that_{\v,\BC}$ itself is an important quantity for the coverage error expansions and feasible inference-optimal bandwidth selectors. This is given by
\begin{equation}
\label{suppeqn:bias rbc}
\E\left[\that_{\v,\BC} \big| X_1, \ldots, X_n \right] - \tau_\v = 
\begin{cases}
O(\h^{\S+\s-\v})  &  \quad \text{ if } \S \leq p+1 \\
\h^{p+2-\v} \mathscr{B}_\BC  +  o_P(\h^{p+2-\v}) & \quad \text{ if } \S\geq p+2,
\end{cases}
\end{equation}
where 
\begin{multline*}
\mathscr{B}_\BC = \frac{\mu_+^{(p+2)}}{(p+2)!}   \v! e_\v'  \Gp{+}^{-1}  \left\{  \L{+}{p,2}  - \rho^{-1}   \Lp{+}  e_{p+1}' \Gq{+}^{-1}    \Lq{+}  \right\}     			\\
-  \frac{\mu_-^{(p+2)}}{(p+2)!}   \v! e_\v'  \Gp{-}^{-1}  \left\{  \L{-}{p,2}  - \rho^{-1}   \Lp{-}  e_{p+1}' \Gq{-}^{-1}    \Lq{-}  \right\},
\end{multline*}
using the notation
\begin{itemize}
	\item $\rho = \h/\b$, 
	\item $\L{+}{p,k} = \Op{+}\left[ X_{\h,1}^{p+k}, \ldots,  X_{\h,n}^{p+k}\right]'/n$, with $\L{+}{p,1} = \Lp{+}$ in particular, and similarly for $\L{-}{p,k}$, and 
	\item $\L{+}{q,k} = \Oq{+}\left[ X_{\b,1}^{q+k}, \ldots,  X_{\b,n}^{q+k}\right]'/n$, with $\L{+}{q,1} = \Lq{+}$ in particular, and similarly for $\L{-}{q,k}$.
\end{itemize}

\begin{remark}[Setting $q > p+1$ or $\rho \to \infty$]
	\label{supprem:higher q}
	It is possible to perform robust bias correction with a polynomial order $q > p+1$ or with $\rho \to\infty$, i.e. a bandwidth $\b$ asymptotically smaller than $\h$. However, neither can improve coverage. The former will tend to inflate variance constants and (to be made feasible) requires estimation of higher derivatives, while the latter leads to a slower variance rate. To see why, first, the general form of $\mathscr{B}_\BC$, provided all derivatives exist (and if they do not, there is even less point to higher $q$) will be
	\begin{align*}
	\mathscr{B}_\BC & = \frac{\mu_+^{(p+2)}}{(p+2)!} \v! e_\v'  \Gp{+}^{-1}  \L{+}{p,2}     - \rho^{-1} \b^{q-p-1}  \frac{\mu_+^{(q+1)}}{(q+1)!}  \nu ! e_\v'  \Gp{+}^{-1} \Lp{+}  e_{p+1}' \Gq{+}^{-1}     \Lq{+}     			\\
	& \qquad  -   \frac{\mu_-^{(p+2)}}{(p+2)!} \v! e_\v'  \Gp{-}^{-1}  \L{-}{p,2}     - \rho^{-1} \b^{q-p-1} \frac{\mu_-^{(q+1)}}{(q+1)!}  \nu ! e_\v'  \Gp{-}^{-1} \Lp{-}  e_{p+1}' \Gq{-}^{-1}     \Lq{-} .
	\end{align*}
	The order of the second term of each line decreases for higher $q$ (provided the same $\h$ sequence is assumed) because the bias of the bias estimator is decreasing. However, the first term of each line, representing the bias not targeted for bias correction, is unchanged. Thus, in rates, nothing can be gained from $q > p+1$. 
	
	Next, suppose that we allow $\rho \to \infty$. Again, the second term in each line is higher order but the first is unchanged, and so the bias rate is not improved (unless $q > p+1$). However, the variance of the estimator will now be determined by $(n\b)^{-1}$ instead of $(n\h)^{-1}$, that is, the variance of the the derivative estimates $\hat{\mu}_+^{(p+1)}$ and $\hat{\mu}_-^{(p+1)}$ is now the dominant variance portion. Setting a finite, positive $\rho$ balances these two.
	
	See \citet*{Calonico-Cattaneo-Farrell_2018_JASA} for further discussion and an expansion with general $q$ in the context of local polynomial regression.
\end{remark}

\subsection{Variance and Variance Estimators}
\label{supp:variance}

To compute the conditional variance define the matrices
\begin{itemize}
	\item $\Sigma_+ = \diag(\sigma^2_+(X_i): i = 1,\ldots, n)$, with $\sigma^2_+(x) = \V[Y(1) \vert X = x]$ defined in Assumption \ref{suppasmpt:srd}, and similarly for $\Sigma_-$.
\end{itemize}

For $\that_\v$, given in \eqref{suppeqn:tau hat}, we find
\begin{align}
\begin{split}
\label{suppeqn:var us}
& \V \left[ \that_\v  \big| X_1, \ldots, X_n \right] = \frac{1}{n\h^{1 + 2\v}}	\mathscr{V},  		\\
& \mathscr{V} = \sp^2 = \frac{\h}{n} \v!^2 \be_\v' \Big( \Gp{+}^{-1} \Op{+} \bS_+ \Op{+}'\Gp{+}^{-1} + \Gp{-}^{-1} \Op{-} \bS_- \Op{-}'  \Gp{-}^{-1} \Big) \be_\v,
\end{split}
\end{align}
where we simultaneously define $\mathscr{V}$ and $\sp^2$. These are identical, but it will frequently be convenient to write $\sp$ rather than $\mathscr{V}^{1/2}$. Compared to the main text, we will often drop the dependence on the bandwidth unless it is required to make a specific point, e.g., we write $\mathscr{V}$ instead of $\mathscr{V}(\h)$.

For $\that_{\v,\BC}$, given in \eqref{suppeqn:that rbc}, we find
\begin{align}
\begin{split}
\label{suppeqn:var rbc}
& \V \left[ \that_{\v,\BC}  \big| X_1, \ldots, X_n \right] = \frac{1}{n\h^{1 + 2\v}}	\mathscr{V}_\BC,  		\\
& \mathscr{V}_\BC = \srbc^2 = \frac{\h}{n} \v!^2 \be_\v' \Big( \Gp{+}^{-1} \Orbc{+} \bS_+ \Orbc{+}'\Gp{+}^{-1} + \Gp{-}^{-1} \Orbc{-} \bS_- \Orbc{-}'  \Gp{-}^{-1} \Big) \be_\v,
\end{split}
\end{align}
where we simultaneously define $\mathscr{V}_\BC$ and $\srbc^2$. These are identical, but it will frequently be convenient to write $\srbc$ rather than $\mathscr{V}_\BC^{1/2}$. Notice that the only change is replacing $\Orbc{+}$ and $\Orbc{-}$ for $\Op{+}$ and $\Op{-}$, as expected from comparing \eqref{suppeqn:that rbc} and \eqref{suppeqn:tau hat}.

To estimate these variances we need only estimate the diagonal matrices $\Sigma_+$ and $\Sigma_-$. Define
\begin{align*}
\Shatp{+} & = \diag\left( ( Y_i - \br_p(X_i - \x)'\bhat_{+,p} )^2: i = 1,\ldots, n\right),  		\\
\Shatp{-} & = \diag\left( ( Y_i - \br_p(X_i - \x)'\bhat_{-,p} )^2: i = 1,\ldots, n\right),  		\\
\Shatrbc{+} & = \diag\left( ( Y_i - \br_q(X_i - \x)'\bhat_{+,q} )^2: i = 1,\ldots, n\right), 
\intertext{and}
\Shatrbc{-} & = \diag\left( ( Y_i - \br_q(X_i - \x)'\bhat_{-,q} )^2: i = 1,\ldots, n\right) ,
\end{align*}
where $\bhat_{+,p}$ and $\bhat_{-,p}$ are given in Equation \eqref{suppeqn:beta hat} and $\bhat_{+,q}$ and $\bhat_{-,q}$ are the same but with $\b$ in place of $\h$ and $q$ in place of $p$.

With these in hand, define
\begin{align}
\begin{split}
\label{suppeqn:se}
\hat{\mathscr{V}} =  \shatp^2 & = \frac{\h}{n} \v!^2 \be_\v' \Big( \Gp{+}^{-1} \Op{+} \Shatp{+} \Op{+}'\Gp{+}^{-1} + \Gp{-}^{-1} \Op{-} \Shatp{-} \Op{-}'  \Gp{-}^{-1} \Big) \be_\v   			\\
\hat{\mathscr{V}}_\BC =  \shatrbc^2 & = \frac{\h}{n} \v!^2 \be_\v' \Big( \Gp{+}^{-1} \Orbc{+} \Shatrbc{+} \Orbc{+}'\Gp{+}^{-1} + \Gp{-}^{-1} \Orbc{-} \Shatrbc{-} \Orbc{-}'  \Gp{-}^{-1} \Big) \be_\v
\end{split}
\end{align}
Other possibilities for standard errors exist, but retaining the fixed-$n$ form is crucial for good coverage properties. For more discussion, including other options and practical details, see \citet*{Calonico-Cattaneo-Farrell_2018_JASA,Calonico-Cattaneo-Farrell_2019_CEOptimal}.

\subsection{Coverage Error Expansion Terms}
\label{supp:terms}

We now give the precise definition of the terms $\mathscr{Q}_{\US,k}$ and $\mathscr{Q}_{\RBC,k}$, $k=1,2,3$, appearing the coverage error expansion in the main text. The final formulas appear at the end of this subsection, and require a fair amount of notation to be defined first. See Section \ref{supp:computing} for the computation of these terms.

We will maintain, as far as possible, fixed-$n$ calculations. All terms must be nonrandom. First, define the following functions, which depend on $n$, $\h$, $\b$, $\v$, $p$, and $K$, though this is mostly suppressed notationally. These functions are all calculated in a fixed-$n$ sense and are all bounded and rateless. 
\begin{align*}
\l^0_{\US}(X_i) & = \v! \be_\v' \left\{\Gpt{+}^{-1} (K_+ \br_p)(\Xhi) - \Gpt{-}^{-1} (K_- \br_p)(\Xhi) \right\} ; 		\\
\l^0_{\RBC}(X_i) & = \l^0_{\US}(X_i)  -   \rho^{p+1} \v! \be_\v' \Gpt{+}^{-1} \Lpt{+} \be_{p+1}' \Gqt{+}^{-1}  (K_+ \br_{p+1})(\Xbi)  		\\
& \qquad   + \rho^{p+1} \v! \be_\v' \Gpt{-}^{-1} \Lpt{-} \be_{p+1}' \Gqt{-}^{-1}  (K_- \br_{p+1})(\Xbi)  ;   		\\
\l^1_{\US}(X_i, X_j) & = \v! \be_\v' \Gpt{+}^{-1} \left( \E[(K_+ \br_p \br_p')(\Xhj)] - (K_+ \br_p \br_p')(\Xhj) \right) \Gpt{+}^{-1} (K_+ \br_p)(\Xhi)  		\\
& \qquad  -  \v! \be_\v' \Gpt{-}^{-1} \left( \E[(K_- \br_p \br_p')(\Xhj)] - (K_- \br_p \br_p')(\Xhj) \right) \Gpt{-}^{-1} (K_- \br_p)(\Xhi)  ; 		\\
\l^1_{\RBC}(X_i, X_j) & = \l^1_{\US}(X_i, X_j)   		\\
& \quad -    \rho^{p+1}   \v! \be_\v' \Gpt{+}^{-1} \biggl\{ \left( \E[(K_+ \br_p \br_p')(\Xhj)] - (K_+ \br_p \br_p')(\Xhj) \right)  \Gpt{+}^{-1}  \Lpt{+} \be_{p+1}'   		\\
& \qquad  +  \left( (K_+ \br_p)(\Xhj) \Xhi^{p+1}   -  \E[(K_+ \br_p)(\Xhj) \Xhi^{p+1}]  \right) \be_{p+1}'       		\\
& \qquad  +  \Lpt{+} \be_{p+1}' \Gqt{+}^{-1}  \left( \E[(K_+ \br_{p+1} \br_{p+1}')(X_{\b,j})] - (K_+ \br_{p+1} \br_{p+1}')(X_{\b,j}) \right)   \biggr\} \Gqt{+}^{-1}  (K_+ \br_{p+1})(\Xbi)  		\\
& \quad  -    \rho^{p+1}   \v! \be_\v' \Gpt{-}^{-1} \biggl\{ \left( \E[(K_- \br_p \br_p')(\Xhj)] - (K_- \br_p \br_p')(\Xhj) \right)  \Gpt{-}^{-1}  \Lpt{-} \be_{p+1}'   		\\
& \qquad  +  \left( (K_- \br_p)(\Xhj) \Xhi^{p+1}   -  \E[(K_- \br_p)(\Xhj) \Xhi^{p+1}]  \right) \be_{p+1}'       		\\
& \qquad  +  \Lpt{-} \be_{p+1}' \Gqt{-}^{-1}  \left( \E[(K_- \br_{p+1} \br_{p+1}')(X_{\b,j})] - (K_- \br_{p+1} \br_{p+1}')(X_{\b,j}) \right)   \biggr\} \Gqt{-}^{-1}  (K_- \br_{p+1})(\Xbi).
\end{align*}
Further, define
\begin{gather*}
\e_i = \One\{X_i < \x\}\e_{-,i}   +  \One\{X_i \geq \x\}\e_{+,i}   		\\
v(X_i) = \One\{X_i < \x\}\sigma^2_-(X_i)   +  \One\{X_i \geq \x\}\sigma^2_+(X_i)  .
\end{gather*}

Let $\i$ (``$\i$'' for Interval) stand in for either $\US$ or $\RBC$.\footnote{More precisely, with this generic ``$\i$'' notation, $\i = \RBC$ refers to quantities appearing in $\mathscr{Q}_{\RBC,k}$, $k=1,2,3$, i.e. those relevant for $\irbc$, which include notations with a subscript $\BC$, such as $\srbc$.} Define $\st_{\i}^2  = \E[ \h^{-1} \l^0_{\i}(X)^2 v(X) ]$.

Now we define three functions $Q_{\US,k}$ and $Q_{\RBC,k}$, $k=1,2,3$ which serve as the main building blocks of the terms of the expansion, capturing in particular all dependence on the data-generating process other than the bias. $Q_{\i,1}$ is the most cumbersome notationally. Begin with the others. Define
\[
Q_{\i,2}(z)  =   -   \st_{\i}^{-2} \left\{ z / 2 \right\}
\]
and
\[
Q_{\i,3}(z)  =   \st_{\i}^{-4} \E [ \h^{-1} \l^0_{\i}(X_i)^3 \e_i^3 ]  \left\{ z^3 / 3 \right\}.
\]

For $\mathscr{Q}_{\i,1}$, it is not quite as simple to state a generic version. Let $\bm{\tilde{G}}_+$ stand in for $\Gpt{+}$ or $\Gqt{+}$ and similarly for $\bm{\tilde{G}}_-$, $\tilde{p}$ stand in for $p$ or $p+1$, and $d_n$ stand in for $\h$ or $\b$, all depending on whether $\ti = \tus$ or $\trbc$. Note however, that $\h$ is still used in many places, in particular for stabilizing fixed-$n$ expectations, for $\trbc$. Indexes $i$, $j$, and $k$ are always distinct (i.e.\ $X_{\h,i} \neq X_{\h,j} \neq X_{\h,k}$). 
\begin{align*}
\hspace{-0.5in} Q_{\i,1}(z)  & =     \st_{\i}^{-6} \E \left[ \h^{-1} \l^0_{\i}(X_i)^3 \e_i^3 \right]^2 \left\{  z^3/3 + 7 z /4 + \st_{\i}^2 z (z^2-3)/4 \right\}   		\\
&  \quad +   \st_{\i}^{-2} \E \left[ \h^{-1} \l^0_{\i}(X_i) \l^1_{\i}(X_i, X_i) \e_i^2 \right] \left\{ - z (z^2 - 3) /2 \right\}   		\\
&  \quad +   \st_{\i}^{-4} \E \left[ \h^{-1} \l^0_{\i}(X_i)^4 (\e_i^4 - v(X_i)^2) \right] \left\{ z(z^2-3)/8 \right\}   		\\
&  \quad -   \st_{\i}^{-2} \E \left[ \h^{-1}\l^0_{\i}(X_i)^2 \br_{\tilde{p}}(X_{d_n,i})'\bm{\tilde{G}}_+^{-1} (K_+ \br_{\tilde{p}})(X_{d_n,i}) \e_i^2 \right] \left\{ z(z^2 - 1)/2 \right\}   		\\
&  \quad -   \st_{\i}^{-2} \E \left[ \h^{-1}\l^0_{\i}(X_i)^2 \br_{\tilde{p}}(X_{d_n,i})'\bm{\tilde{G}}_-^{-1} (K_- \br_{\tilde{p}})(X_{d_n,i}) \e_i^2 \right] \left\{ z(z^2 - 1)/2 \right\}   		\\
&  \quad -   \st_{\i}^{-4} \E \left[ \h^{-1} \l^0_{\i} (X_i)^3  \br_{\tilde{p}}(X_{d_n,i})'\bm{\tilde{G}}_+^{-1} \e_i^2 \right] \E \left[ \h^{-1} (K_+ \br_{\tilde{p}})(X_{d_n,i}) \l^0_{\i} (X_i) \e_i^2 \right] \left\{ z(z^2 - 1)  \right\}   		\\
&  \quad -   \st_{\i}^{-4} \E \left[ \h^{-1} \l^0_{\i} (X_i)^3  \br_{\tilde{p}}(X_{d_n,i})'\bm{\tilde{G}}_-^{-1} \e_i^2 \right] \E \left[ \h^{-1} (K_- \br_{\tilde{p}})(X_{d_n,i}) \l^0_{\i} (X_i) \e_i^2 \right] \left\{ z(z^2 - 1)  \right\}   		\\
&  \quad +   \st_{\i}^{-2} \E \left[ \h^{-2} \l^0_{\i}(X_i)^2 (\br_{\tilde{p}}(X_{d_n,i})'\bm{\tilde{G}}_+^{-1}(K_+ \br_{\tilde{p}})(X_{d_n,j}) )^2 \e_j^2 \right] \left\{ z(z^2 - 1)/4 \right\}   		\\
&  \quad +   \st_{\i}^{-2} \E \left[ \h^{-2} \l^0_{\i}(X_i)^2 (\br_{\tilde{p}}(X_{d_n,i})'\bm{\tilde{G}}_-^{-1}(K_- \br_{\tilde{p}})(X_{d_n,j}) )^2 \e_j^2 \right] \left\{ z(z^2 - 1)/4 \right\}   		\\
&  \quad +   \st_{\i}^{-4} \E \left[ \h^{-3} \l^0_{\i} (X_j)^2 \br_{\tilde{p}}(X_{d_n,j})'\bm{\tilde{G}}_+^{-1}(K_+ \br_{\tilde{p}})(X_{d_n,i}) \l^0_{\i}(X_i) \br_{\tilde{p}}(X_{d_n,j})'\bm{\tilde{G}}_+^{-1}(K_+ \br_{\tilde{p}})(X_{d_n,k}) \l^0_{\i}(X_k)  \e_i^2 \e_k^2 \right]   		\\
&  \qquad \qquad \qquad \qquad \qquad   \times \;   \left\{  z(z^2 - 1) /2 \right\}   		\\
&  \quad +   \st_{\i}^{-4} \E \left[ \h^{-3} \l^0_{\i} (X_j)^2 \br_{\tilde{p}}(X_{d_n,j})'\bm{\tilde{G}}_-^{-1}(K_- \br_{\tilde{p}})(X_{d_n,i}) \l^0_{\i}(X_i) \br_{\tilde{p}}(X_{d_n,j})'\bm{\tilde{G}}_-^{-1}(K_- \br_{\tilde{p}})(X_{d_n,k}) \l^0_{\i}(X_k)  \e_i^2 \e_k^2 \right]   		\\
&  \qquad \qquad \qquad \qquad \qquad   \times \;   \left\{  z(z^2 - 1) /2 \right\}   		\\
&  \quad +   \st_{\i}^{-4} \E \left[ \h^{-1} \l^0_{\i}(X_i)^4 \e_i^4 \right] \left\{ - z (z^2 - 3)/24 \right\}   		\\
&  \quad +   \st_{\i}^{-4} \E \left[ \h^{-1} \left( \l^0_{\i}(X_i)^2 v(X_i) - \E[\l^0_{\i}(X_i)^2 v(X_i)] \right) \l^0_{\i}(X_i)^2 \e_i^2 \right] \left\{ z(z^2 - 1)/4 \right\}   		\\
&  \quad +   \st_{\i}^{-4} \E \left[ \h^{-2} \l^1_{\i}(X_i, X_j) \l^0_{\i}(X_i)\l^0_{\i}(X_j)^2 \e_j^2 v(X_i) \right] \left\{  z (z^2 - 3) \right\}   		\\
&  \quad +   \st_{\i}^{-4} \E \left[ \h^{-2} \l^1_{\i}(X_i, X_j)  \l^0_{\i}(X_i) \left( \l^0_{\i}(X_j)^2 v(X_j) - \E[\l^0_{\i}(X_j)^2 v(X_j)] \right) \e_i^2 \right] \left\{ - z \right\}   		\\
&  \quad +   \st_{\i}^{-4} \E \left[ \h^{-1}  \left( \l^0_{\i}(X_i)^2 v(X_i) - \E[\l^0_{\i}(X_i)^2 v(X_i)] \right)^2 \right] \left\{ - z(z^2 + 1) /8 \right\}   .
\end{align*}

For computation, note that the tenth and eleventh terms can be rewritten by factoring the expectation, after rearranging the terms using the fact that $\br_{\tilde{p}}(X_{d_n,j})'\bm{\tilde{G}}^{-1}\br_{\tilde{p}}(X_{d_n,i})$ is a scalar, as follows:
\begin{align*}
& \E \left[ \h^{-3} \l^0_{\i} (X_j)^2 \br_{\tilde{p}}(X_{d_n,j})'\bm{\tilde{G}}^{-1}(K \br_{\tilde{p}})(X_{d_n,i}) \l^0_{\i}(X_i) \br_{\tilde{p}}(X_{d_n,j})'\bm{\tilde{G}}^{-1}(K \br_{\tilde{p}})(X_{d_n,k}) \l^0_{\i}(X_k)  \e_i^2 \e_k^2 \right]   		\\
& \qquad = \E\left[\h^{-1} \l^0_{\i}(X_i) \e_i^2 (K \br_{\tilde{p}}')(X_{d_n,i}) \bm{\tilde{G}}^{-1} \right] \; \E\left[\h^{-1} \br_{\tilde{p}}(X_{d_n,j}) \l^0_{\i} (X_j)^2 \br_{\tilde{p}}(X_{d_n,j})'\bm{\tilde{G}}^{-1}  \right]    		\\
& \qquad\qquad\qquad\qquad      \times \; \E\left[\h^{-1} (K \br_{\tilde{p}})(X_{d_n,k})   \l^0_{\i}(X_k)  \e_k^2  \right].
\end{align*}
This will greatly ease implementation.

The final ingredient required to define the $\mathscr{Q}_{\US,k}$ and $\mathscr{Q}_{\RBC,k}$ terms is the bias. The expressions in Equations \eqref{suppeqn:bias us} and \eqref{suppeqn:bias rbc} can not be used as these are random. Instead, their fixed-$n$ analogues will appear. To this end, define
\[ \tilde{\mathscr{B}}_\US  = \frac{\v!}{(p+1)!} \be_\v' \left( \Gpt{+}^{-1} \Lpt{+} \mu_+^{(p+1)} - \Gpt{-}^{-1} \Lpt{-} \mu_-^{(p+1)}\right) \]
and
\begin{multline*}
\tilde{\mathscr{B}}_\BC = \frac{\mu_+^{(p+2)}}{(p+2)!}   \v! e_\v'  \Gpt{+}^{-1}  \left\{  \Lt{+}{p,2}  - \rho^{-1}   \Lpt{+}  e_{p+1}' \Gqt{+}^{-1}    \Lqt{+}  \right\}     			\\
-  \frac{\mu_-^{(p+2)}}{(p+2)!}   \v! e_\v'  \Gpt{-}^{-1}  \left\{  \Lt{-}{p,2}  - \rho^{-1}   \Lpt{-}  e_{p+1}' \Gqt{-}^{-1}    \Lqt{-}  \right\},
\end{multline*}
where
\begin{itemize}
	\item $\Gpt{+} = \E[\Gp{+}]$, $\Lpt{+} = \E[\Lp{+}]$, and so forth.
\end{itemize}

Finally, the $\mathscr{Q}_{\US,k}$ and $\mathscr{Q}_{\RBC,k}$ terms are defined as follows, where as usual $\i$ stands in for either $\ius$ or $\irbc$, 
\begin{align}
\begin{split}
\label{suppeqn:terms}
\mathscr{Q}_{\i,1} & = 2 \phi(z_{\alpha/2}) Q_{\i,1}(z_{\alpha/2})   		\\
\mathscr{Q}_{\i,2} & = 2 \phi(z_{\alpha/2}) Q_{\i,2}(z_{\alpha/2}) \tilde{\mathscr{B}}_\i^2   		\\
\mathscr{Q}_{\i,3} & = 2 \phi(z_{\alpha/2}) Q_{\i,3}(z_{\alpha/2}) \tilde{\mathscr{B}}_\i   		\\
\end{split}
\end{align}

\section{Main Results: Coverage Error and Edgeworth Expansions}
\label{supp:theorems}

We now state the main theoretical results: Edgeworth expansion for the distributions of the $t$-statistics 
\begin{equation}
\label{suppeqn:t stats}
\tus = \frac{\sqrt{n\h^{1+2\v}}(\that_{\v,\US} - \tau_\v)}{\shatus}   		\qquad \text{ and }\qquad   		  \trbc = \frac{\sqrt{n\h^{1+2\v}}(\that_{\v,\BC} - \tau_\v)}{\shatrbc}.
\end{equation}
The point estimators are given in Equations \eqref{suppeqn:tau hat} and \eqref{suppeqn:that rbc} and the standard errors are in \eqref{suppeqn:se}.

Before stating the results, more notation is needed. In addition to the terms $Q_{\i,k}$, $k=1,2,3$, two other terms appear in the Edgeworth expansion for the $t$-statistic, which then cancel upon computing coverage error due to symmetry. These are
\begin{align*}
Q_{\i,4}(z)   =    \st_{\i}^{-3} \E \left[ \h^{-1} \l^0_{\i}(X_i)^3 \e_i^3 \right] \left\{  (2z^2 - 1)/6 \right\}  		  \qquad \text{ and } \qquad  		  Q_{\i,5}(z)   =   -   \st_{\i}^{-1}.
\end{align*}

The coverage error expansions follow immediately from the results below by taking the difference of expansions at $z_{1-\alpha/2}$ and $z_{\alpha/2}$. It is clearest to state separate results for $\tus$ and $\trbc$. For the standard, or undersmoothing, approach, we have the following result.

\begin{theorem}[Edgeworth Expansion for $\tus$]
	\label{suppthm:ee us}
	Suppose Assumptions \ref{suppasmpt:srd} and \ref{suppasmpt:kernel} hold with $\S \geq p+1$. If $n \h/ \log(n\h)^{2+\gamma} \to \infty$ and $\sqrt{n\h} \h^{p+1} \log(n\h)^{1+\gamma} \to 0$, for any $\gamma>0$, then 	
	\[ 
	\sup_{z\in\R} \left| \P[\tus < z ] - \Phi(z) - \phi(z) \mathcal{E}_\US(z) \right| = \epsilon_\US,
	\]
	where $\epsilon_\US = o((n\h)^{-1}) + O(n \h^{3 + 2p + 2\s} + \h^{p+1 + \s})$ and
	\[ \mathcal{E}_\US(z) =  \frac{1}{n \h}  Q_{\ius,1}  +   n \h^{3 + 2p}  Q_{\ius,2} \tilde{\mathscr{B}}_\US^2   +   \h^{p+1}   Q_{\ius,3} \tilde{\mathscr{B}}_\US  +  \frac{1}{\sqrt{n \h}}  Q_{\ius,4}(z)  +  \sqrt{n\h} \h^{p+1} \tilde{\mathscr{B}}_\US Q_{5,\ius}(z). \]
\end{theorem}
This immediately yields the follow result for optimal undersmoothing, analogous to the result for robust bias correction in the paper. 
\begin{corollary}
	\label{suppthm:optimal us}
	Let the conditions of Theorem \ref{suppthm:ee us} hold. Then the fastest coverage error decay possible is $\P[\tau_\nu \in I_\US(h)] = (1 - \alpha) + O\left(n^{-(p+1)/(p+2)}\right)$ and is attained by choosing $\h \asymp n^{-1/(p+2)}$. In particular, if $\mathscr{Q}_{\US,k} \neq 0, k=1,2,3$, the optimal bandwidth is given by
	\[ 
	\h_\US = \mathcal{H}_\US n^{-1/(p+2)},  		\qquad \text{ with } \qquad 		  \mathcal{H}_\US = \argmin_{H>0} \left| \frac{1}{H} \mathscr{Q}_{\ius,1}   +   H^{3+2p} \mathscr{Q}_{\ius,2} + H^{1+p} \mathscr{Q}_{\ius,3} \right|.
	\]
\end{corollary}

Turning to robust bias correction, we differentiate between the case when $\S \geq p+2$, allowing all bias terms to be characterized, and the case when there is not sufficient smoothness to do so. 

\begin{theorem}[Edgeworth Expansions for $\trbc$]
	\label{suppthm:ee rbc}
	Suppose Assumptions \ref{suppasmpt:srd} and \ref{suppasmpt:kernel} hold. Assume $n \h/ \log(n\h)^{2+\gamma} \to \infty$ for any $\gamma>0$ and $\rho=h/b \to \bar{\rho} < \infty$.
	\begin{enumerate}
		\item If $\S \geq p+2$ and $\sqrt{n\h} \h^{p+2}(1 + \rho^{-1}) \log(n\h)^{1+\gamma} \to 0$ for any $\gamma>0$ then 	
		\[ 
		\sup_{z\in\R} \left| \P[\trbc < z ] - \Phi(z) - \phi(z) \mathcal{E}_\RBC(z) \right| =  \epsilon_\RBC,\]
		where $\epsilon_\RBC = o((n\h)^{-1}) + O(n \h^{5 + 2p + 2\s} + \h^{p+2 + \s})$ and
		\begin{multline*}
		\mathcal{E}_\RBC(z) = \frac{1}{n \h} \phi(z) Q_{\irbc,1}  +   n \h^{5 + 2p} \phi(z) Q_{\irbc,2} \tilde{\mathscr{B}}_\BC^2  +   \h^{p+2}  \phi(z) Q_{\irbc,3} \tilde{\mathscr{B}}_\BC        		\\
		+   \frac{1}{\sqrt{n \h}}  Q_{\irbc,4}(z)  +  \sqrt{n\h} \h^{p+1} \tilde{\mathscr{B}}_\BC Q_{5,\irbc}(z).
		\end{multline*}
		
		\item If $\S \geq p+1$ and $\sqrt{n\h} \h^{p+1}(1 + \rho^{-1}) \log(n\h)^{1+\gamma} \to 0$ for any $\gamma>0$ then 	
		\[ 
		\sup_{z\in\R} \left| \P[\trbc < z ] - \Phi(z) -  \frac{1}{n \h} \phi(z) Q_{\irbc,1}  -  \frac{1}{\sqrt{n \h}} \phi(z) Q_{\irbc,4}(z)  -   \btrbc Q_{5,\irbc}(z)    \right| =  \epsilon_\RBC,\]
		where $\epsilon_\RBC = o((n\h)^{-1}) + O(n \h^{3 + 2p + 2\s} + \h^{p+1 + \s})$ and
		\begin{align*}
		\btrbc & = \sqrt{n\h} \;  \v! \be_\v'\Gpt{+}^{-1} \E\biggl[ \Bigl\{ \h^{-1} (K_+ \br_p)(\Xhi)    -    \rho^{p+1} \Lpt{+} \be_{p+1}'\Gqt{+}^{-1} \b^{-1} (K_+ \br_{p+1})(\Xbi) \Bigr\}   	   			\\
		& \qquad\qquad\qquad  \times   \left( \mu_+(X_i) - \br_{p+1}(X_i - \x)'\bbeta_{+,p+1}\right)   \biggr]   			\\
		& \quad  -  \sqrt{n\h} \;  \v! \be_\v'\Gpt{-}^{-1} \E\biggl[ \Bigl\{ \h^{-1} (K_- \br_p)(\Xhi)    -    \rho^{p+1} \Lpt{-} \be_{p+1}'\Gqt{-}^{-1} \b^{-1} (K_- \br_{p+1})(\Xbi) \Bigr\}   	   			\\
		& \quad \qquad\qquad\qquad  \times   \left( \mu_-(X_i) - \br_{p+1}(X_i - \x)'\bbeta_{-,p+1}\right)   \biggr]   ,			
		\end{align*}
		with $\bbeta_{+,k}$ the $k+1$ vector with $(j+1)$ element equal to $\mu_+^{(j)}(\x)/j!$ for $j = 0, 1, \ldots, k$ as long as $j \leq S$, and zero otherwise, and similarly for $\bbeta_{-,k}$.
	\end{enumerate}
	
\end{theorem}

\section{Proofs for Main Results}
\label{supp:proofs}


We now present proofs for the main theoretical results. We present details for Theorem \ref{suppthm:ee us}, as the proof for Theorem \ref{suppthm:ee rbc} is largely similar; a brief discussion is given. We first prove the validity of the expansion, deferring computation of the terms to a subsection below. We will rely on some technical results from the supplement to \citet*{Calonico-Cattaneo-Farrell_2019_CEOptimal}, which in general contains more detailed proofs, though in the context of nonparametric regression rather than RD.

The first step in the proof is to show that 
\begin{equation}
\label{suppeqn:step 1}
\P \left[ \tus < z \right] = \P \left[ \breve{T} < z \right]  + o\left( (n\h)^{-1} + \h^{p+1}  +  n \h^{3 + 2p} \right),
\end{equation}
for a smooth function $\breve{T} := \breve{T}((n\h)^{-1/2} \sumi \bZ_i)$, where $\bZ_i$ a random vector consisting of functions of the data, that, among other requirements, obeys Cram\'er's condition under our assumptions.

Define
\begin{itemize}
	\item $\tO = \sqrt{n\h}$.
\end{itemize}
The $t$-statistic at hand is
\[
\tus = \frac{\sqrt{n\h^{1+2\v}}(\that_{\v,\US} - \tau_\v)}{\shatus} = \frac{ \tO \v!\be_\nu'\left( \Gp{+}^{-1} \Op{+}  \left( \bY  - \bR \bbeta_{+,p} \right)  -  \Gp{-}^{-1} \Op{-} \left( \bY  - \bR \bbeta_{-,p} \right) \right)/n }{\shatus}.
\]
The numerator is already a smooth function of well-behaved random variables (obeying Cram\'er's condition in particular), therefore the difference between $\tus$ and $\breve{T}$ lies in the denominator. Recall from \eqref{suppeqn:se} that
\[
\hat{\mathscr{V}} =  \shatp^2  = \frac{\h}{n} \v!^2 \be_\v' \Big( \Gp{+}^{-1} \Op{+} \Shatp{+} \Op{+}'\Gp{+}^{-1} + \Gp{-}^{-1} \Op{-} \Shatp{-} \Op{-}'  \Gp{-}^{-1} \Big) \be_\v.
\]
As with the numerator, the $\G{\bullet}{p}$ matrices are already in the appropriate form. We must expand the ``meat'' portions, $\h \Op{+} \Shatp{+} \Op{+}'/n$ and $\h \Op{-} \Shatp{-} \Op{-}'/n$, and their estimated residuals. The expansions for the two, being additive, can be done separately. We state only the ``$+$'' terms. Let $\e_{+,i} = Y_i(1)  - \mu_+(X_i)$. Then expand
\begin{align}
\begin{split}
\label{suppeqn:var expansion}
\frac{\h}{n} \Op{+} \Shatp{+} \Op{+}' & = \frac{1}{n \h} \sumi (K_+^2 \br_p \br_p')(\Xhi) \left(  Y_i - \br_p(X_i - \x)'\bhat_+ \right)^2   		 \\
& = \frac{1}{n \h} \sumi (K_+^2 \br_p \br_p')(\Xhi)  \Big( \e_i  + \left[\mu_+(X_i) -  \br_p(X_i - \x)'\bbeta_{+,p} \right]     			\\
& \qquad\qquad\qquad\qquad \qquad\qquad   +  \br_p(X_i - \x)'\left[ \bbeta_{+,p} - \bhat_+\right] \Big)^2 	 \\
& = \frac{1}{n \h} \sumi (K_+^2 \br_p \br_p')(\Xhi)  \Big( \e_i  + \left[\mu_+(X_i) -  \br_p(X_i - \x)'\bbeta_{+,p}\right]    		\\
& \qquad\qquad\qquad\qquad \qquad\qquad   -  \br_p(\Xhi)' \Gp{+}^{-1} \Op{+} \left[\bY - \bR \bbeta_{+,p}\right] /n \Big)^2. 	
\end{split}
\end{align}
Define
\begin{align*}
V^+_1 & = \frac{1}{n \h} \sumi (K_+^2 \br_p \br_p')(\Xhi) \e_i^2,   		\\  
V^+_2 & = \frac{1}{n \h} \sumi (K_+^2 \br_p \br_p' \br_p')(\Xhi) \e_i \Gp{+}^{-1} \Op{+} \left[\bY - \bR \bbeta_{+,p}\right] /n,  		\\   
V^+_3 & = \frac{1}{n \h} \sumi (K_+^2 \br_p \br_p')(\Xhi)  \left[\mu_+(X_i) -  \br_p(X_i - \x)'\bbeta_{+,p}\right]^2   ,			\\    
V^+_4 & = \frac{1}{n \h} \sumi (K_+^2 \br_p \br_p')(\Xhi) \left\{  \e_i \left[\mu_+(X_i) -  \br_p(X_i - \x)'\bbeta_{+,p}\right] \right\},  		\\  
V^+_5 & = \frac{1}{n \h} \sumi (K_+^2 \br_p \br_p' \br_p')(\Xhi) \left[\mu_+(X_i) -  \br_p(X_i - \x)'\bbeta_{+,p}\right] \Gp{+}^{-1} \Op{+} \left[\bY - \bR \bbeta_{+,p}\right]/n ,  		\\  
V^+_6 & = \frac{1}{n \h} \sumi (K_+^2 \br_p \br_p')(\Xhi)  \big\{ \br_p(\Xhi)'\Gp{+}^{-1} \Op{+} \left[\bY - \bR \bbeta_{+,p}\right] /n \big\}^2,      		 
\end{align*}
and
\begin{align*}
\breve{V}^+_5  &  =   \sum_{l_i=0}^p \sum_{l_j=0}^p  \left[\Gp{+}^{-1}\right]_{l_i, l_j} \E\left[ (K_+^2 \br_p \br_p') (\Xhi) (\Xhi)^{l_i} \left(\mu_+(X_i) -  \br_p(X_i - \x)'\bbeta_{+,p}\right)  \right]   			 \\
& \qquad \qquad \qquad\qquad    \times   \frac{1}{n \h} \sumj \bigg\{  K_+(\Xhj)  (\Xhj)^{l_j}  \left(Y_j - \br_p(X_j - \x)'\bbeta_{+,p}\right)  \bigg\},    			\\
\breve{V}^+_6 & = \sum_{l_{i_1}=0}^p \sum_{l_{i_2}=0}^p \sum_{l_{j_1}=0}^p \sum_{l_{j_2}=0}^p   \left[\Gp{+}^{-1}\right]_{l_{i_1}, l_{j_1}}  \left[\Gp{+}^{-1}\right]_{l_{i_2}, l_{j_2}}  \E\left[ \h^{-1} (K_+^2 \br_p \br_p')(\Xhi)   (\Xhi)^{l_{i_1} + l_{i_2}}   \right]		  			 \\
& \quad  \times  
\frac{1}{(n \h)^2}   \sumj \sumk    K_+(\Xhj) (\Xhj)^{l_{j_1}}  \left(Y_j - \br_p(X_j - \x)'\bbeta_{+,p}\right)  K_+(\Xhk)   (\Xhk)^{l_{j_2}}  \left(Y_k - \br_p(X_k - \x)'\bbeta_{+,p}\right) .
\end{align*}

where $\left[\Gp{+}^{-1}\right]_{l_i, l_j}$ is the $\{l_i+1, l_j+1\}$ element of $\Gp{+}^{-1}$, and similarly define the corresponding ``$-$'' versions of all these. 

With these definitions in hand, rewrite $\hat{\mathscr{V}} =  \shatp^2$ as 
\begin{multline*}
\label{suppeqn:shat terms}
\shatus^2 = \v!^2 \be_\v'\Gp{+}^{-1}  \Big( V^+_1  +  2 V^+_4  - 2 V^+_2  + V^+_3  - 2 V^+_5  + V^+_6 \Big) \Gp{+}^{-1} \be_\v   			\\   
+ \v!^2 \be_\v'\Gp{-}^{-1}  \Big( V^-_1  +  2 V^-_4  - 2 V^-_2  + V^-_3  - 2 V^-_5  + V^-_6 \Big) \Gp{-}^{-1} \be_\v
\end{multline*}
and let
\begin{multline*}
\sbp^2 = \v!^2 \be_\v'\Gp{+}^{-1}  \Big( V^+_1 -  2 V^+_2  +  2 V^+_4  -  2 \breve{V}_5  + \breve{V}_6 \Big) \Gp{+}^{-1} \be_\v   		\\
+  \v!^2 \be_\v'\Gp{-}^{-1}  \Big( V^-_1 -  2 V^-_2  +  2 V^-_4  -  2 \breve{V}^-_5  + \breve{V}^-_6 \Big) \Gp{-}^{-1} \be_\v.
\end{multline*}
Then, referring back to Equation \eqref{suppeqn:step 1}, we have 
\[
\P \left[ \tp < z \right] = \P \left[ \breve{T}  +  U_n < z \right],
\]
with 
\begin{equation*}
\label{suppeqn:U}
U_n = \left( \shatp^{-1} -  \sbp^{-1}\right) \tO \v!\be_\nu'\left( \Gp{+}^{-1} \Op{+}  \left( \bY  - \bR \bbeta_{+,p} \right)  -  \Gp{-}^{-1} \Op{-} \left( \bY  - \bR \bbeta_{-,p} \right) \right)/n
\end{equation*}
and
\begin{align*}
\breve{T} & =  \sbp^{-1} \tO \v!\be_\nu'\left( \Gp{+}^{-1} \Op{+}  \left( \bY  - \bR \bbeta_{+,p} \right)  -  \Gp{-}^{-1} \Op{-} \left( \bY  - \bR \bbeta_{-,p} \right) \right)/n.
\end{align*}
As required, $\breve{T}:= \breve{T}(\tO^{-1} \sumi \bZ_i)$ is a smooth function of the sample average of $\bZ_i = ({\bZ^+_i}',{\bZ^-_i}')'$, where $\bZ^+_i$ is defined as
\begin{align*}
\begin{split}
\label{suppeqn:bZ}
\bZ^+_i  = \Bigg(
& \Big\{  (K_+ \br_p)(\Xhi) (Y_i - \br_p(X_i - \x)'\bbeta_{+,p})  \Big\}'  ,    		\\ 		  
& \vech\Big\{  (K_+ \br_p \br_p')(\Xhi)  \Big\}' ,    		\\ 		  
& \vech\Big\{  (K_+^2 \br_p \br_p')(\Xhi) \e_{+,i}^2   \Big\}'  ,    		\\ 		  
& \vech\Big\{  (K_+^2 \br_p \br_p' )(\Xhi) (\Xhi)^0 \e_{+,i}   \Big\}', \vech\Big\{  (K_+^2 \br_p \br_p' )(\Xhi) (\Xhi)^1 \e_{+,i}   \Big\}',   		\\
& \qquad \vech\Big\{  (K_+^2 \br_p \br_p' )(\Xhi) (\Xhi)^2 \e_{+,i}  \Big\}' ,\ldots, \vech\Big\{  (K_+^2 \br_p \br_p' )(\Xhi) (\Xhi)^p \e_{+,i}   \Big\}' ,	\\   
& \vech\Big\{  (K_+^2 \br_p \br_p')(\Xhi) \big\{  \e_{+,i} \big[\mu(X_i) -  \br_p(X_i - \x)'\bbeta_{+,p}\big] \big\}   \Big\}'   		  
\Bigg)',
\end{split}
\end{align*}
and $\bZ^-_i$ is analogous. In order of their listing above, these pieces come from (i) the ``score'' portion of the numerator, (ii) the ``Gram'' matrix $\Gp{+}$, (iii) $V^+_1$, (iv) $V^+_2$, and (v) $V^+_4$. Notice that $\breve{V}^+_5$ and $\breve{V}^+_6$ do not add any additional elements to $\bZ_i$.

Equation \eqref{suppeqn:step 1} now follows from the Delta method for Edgeworth expansions (see Lemma S.II.1 of \citealp*{Calonico-Cattaneo-Farrell_2019_CEOptimal} and discussion there) if we can show that
\begin{equation}
\label{suppeqn:delta}
r_{\ip}^{-1} \P [|U_n| > r_n] = o(1),
\end{equation}
where $r_{\ip} = \max\{\tO^{-2}, n \h^{3+2p}, \h^{p+1} \}$ and $r_n = o(r_{\ip})$.

For a point $\bar{\seletter}^2 \in [\sbus^2, \shatus^2]$, a Taylor expansion gives
\begin{equation*}
\label{suppeqn:variance taylor}
\shatus^{-1} - \sbus^{-1} =  - \frac{1}{2} \frac{ \shatus^2 - \sbp^2}{\sbp^3}  + \frac{3}{8} \frac{  \left( \shatus^2 - \sbus^2\right)^2}{\bar{\seletter}^5} .
\end{equation*}
Therefore, if $\left| \shatp^2 - \sbp^2 \right| = o_p(1)$, the result in \eqref{suppeqn:delta} will hold once we have shown that
\begin{align*}
r_{\ip}^{-1} & \P \left[\left| \left( \shatp^2 - \sbp^2\right) \left(\tO \v!\be_\nu'\left( \Gp{+}^{-1} \Op{+}  \left( \bY  - \bR \bbeta_{+,p} \right)  -  \Gp{-}^{-1} \Op{-} \left( \bY  - \bR \bbeta_{-,p} \right) \right)/n \right) \right|  > r_n  \right]    		\\
& = r_{\ip}^{-1} \P \Bigg[\Bigg|  \left(  \v!^2 \be_\v'\Gp{+}^{-1}  \left( V^+_3   -  2 [V^+_5 - \breve{V}^+_5]  +   [V^+_6 - \breve{V}^+_6]  \right) \Gp{+}^{-1} \be_\v \right)    			\\	
& \qquad \qquad\quad    \times   \left(  \v!^2 \be_\v'\Gp{-}^{-1}  \left( V^-_3   -  2 [V^-_5 - \breve{V}^-_5]  +   [V^-_6 - \breve{V}^-_6]  \right) \Gp{-}^{-1} \be_\v \right)    			\\	
& \qquad \qquad\quad   \times \Big( \tO \v!\be_\nu'\left( \Gp{+}^{-1} \Op{+}  \left( \bY  - \bR \bbeta_{+,p} \right)  -  \Gp{-}^{-1} \Op{-} \left( \bY  - \bR \bbeta_{-,p} \right) \right)/n \Big)
\Bigg|  > r_n  \Bigg]   		 \\
& = o(1). 
\end{align*}
Recall that $r_{\ip} = \max\{\tO^{-2}, n \h^{3+2p}, \h^{p+1} \}$ and $r_n = o(r_{\ip})$. The result then follows by the same argument as Section S.II.5.1 of \citet*{Calonico-Cattaneo-Farrell_2019_CEOptimal}; cf.\ their Equation (S.II.23) and notice that all products of ``$+$'' and ``$-$'' are zero because of their respective indicator functions.

Thus we have established Equation \eqref{suppeqn:step 1}. Section S.II.5.2 of \citet*{Calonico-Cattaneo-Farrell_2019_CEOptimal} shows that $\sumi \V[\bZ_i]^{-1/2}(\bZ_i - \E[\bZ_i])/\sqrt{n}$ obeys an Edgeworth expansion. From this, we deduce that $\breve{T} = \breve{T}\left( \V[\bZ_i]^{1/2} S_n + n \E[\bZ_i]/ \tO \right)$ has its own expansion by \citet*{Skovgaard1986_ISR}, and the result for $\tus$ holds by combining the expansion for $\breve{T}$ with Equation \eqref{suppeqn:step 1}. This completes the proof of Theorem \ref{suppthm:ee us}. \qed

Let us turn to Theorem \ref{suppthm:ee rbc}. The starting point of the proof is the same as that of Theorem \ref{suppthm:ee us}: the $t$-statistic. Looking at the two $t$-statistics in \eqref{suppeqn:t stats}, and the definitions of the respective point estimators, \eqref{suppeqn:tau hat} and \eqref{suppeqn:that rbc}, and standard errors, \eqref{suppeqn:se}, we see that the only substantive differences are the matrices $\O{\pm}{\bullet}$. The estimated residuals are of the same form as above, with only the bandwidth and polynomial order changed. These changes are reflected in the expansion already. The key is thus to redo the expansion of \eqref{suppeqn:var expansion} with $\Orbc{\pm}$ in place of $\Op{\pm}$. The latter lead to the weights $(K_+^2 \br_p \br_p')(\Xhi)$, and these are simply replaced by
\[ \Big((K_+ \br_p)(\Xhi) -  \rho^{p+1}  \Lp{+} \be_{p+1}' \Gq{+}^{-1} (K_+ \br_{p+1})(\Xbi)\Big)\Big((K_+ \br_p)(\Xhi) -  \rho^{p+1}  \Lp{+} \be_{p+1}' \Gq{+}^{-1} (K_+ \br_{p+1})(\Xbi)\Big)'. \]
The same steps are then repeated and hold exactly as before, with the corresponding changes to the rates and terms of the expansion. These are all built into the notation. For more details, see Section S.II.6 of \citet*{Calonico-Cattaneo-Farrell_2019_CEOptimal}. \qed

\subsection{Computing the Terms of the Expansion}
\label{supp:computing}

Computing the terms of the Edgeworth expansions of Theorems \ref{suppthm:ee us} and \ref{suppthm:ee rbc}, listed in Section \ref{supp:terms}, is straightforward but tedious. We give a short summary here, following the essential steps of \cite[Chapter 2]{Hall1992_book} and \citet*{Calonico-Cattaneo-Farrell_2019_CEOptimal}. In what follows, will always discard higher order terms and write $A \oeq B$ to denote $A = B + o((n\h)^{-1} + \h^{p+1}  +  n \h^{3 + 2p})$. 

We will need much of the notation defined in Section \ref{supp:terms}. As there, let $\bm{\tilde{G}}_+$ stand in for $\Gpt{+}$ or $\Gqt{+}$ and similarly for $\bm{\tilde{G}}_-$, $\tilde{p}$ stand in for $p$ or $p+1$, and $d_n$ stand in for $\h$ or $\b$, all depending on if $\ti = \tus$ or $\trbc$. Note however, that $\h$ is still used in many places, in particular for stabilizing fixed-$n$ expectations, for $\trbc$. Indexes $i$, $j$, and $k$ are always distinct (i.e.\ $X_{\h,i} \neq X_{\h,j} \neq X_{\h,k}$).

The steps to compute the expansion are as follows. First, we compute a Taylor expansion of $\ti$ around nonrandom denominators, including both $\shat^{-1}$ and $\bm{\tilde{G}}^{-1}$. The cumulants of this linearized version are the approximate cumulants of $\ti$ itself, which determine the terms of the expansion, as described by \citet*{Bhattacharya-Rao1976_book} and \citet*{Hall1992_book}. 

It is important to note that the functions $\l^0_{\i}(X_i)$ and $\l^1_{\i}(X_i, X_j)$ already include terms to the left and right of the cutoff. The same is true of
\begin{gather*}
\e_i = \One\{X_i < \x\}\e_{-,i}   +  \One\{X_i \geq \x\}\e_{+,i}   		\\
v(X_i) = \One\{X_i < \x\}\sigma^2_-(X_i)   +  \One\{X_i \geq \x\}\sigma^2_+(X_i)  .
\end{gather*}
Notice that, because of the indicator functions for each side, products such as $\l^0_{\i}(X_i)^2$ or $\l^0_{\i}(X_i)\l^1_{\i}(X_i, X_j)$ or $\l^0_{\i}(X_i)\e_i^2$, etc., are always correct.

The Taylor expansion is
\begin{multline*}
\ti \oeq \left\{ 1 - \frac{1}{2 \st_{\i}^2}\left( W_{\i,1}  +  W_{\i,2}  + W_{\i,3}\right) + \frac{3}{8 \st_{\i}^4} \left( W_{\i,1}  +  W_{\i,2}  + W_{\i,3}\right)^2  \right\}  		\\ 		\times \;  \st_{\i}^{-1} \left\{   E_{\i,1} + E_{\i,2} + E_{\i,3} + B_{\i,1} \right\},
\end{multline*}
where
\begin{align*}
W_{\i,1} & =  \frac{1}{n\h} \sumi \left\{\l^0_{\i}(X_i)^2  \left( \e_i^2 - v(X_i) \right) \right\}     			\\
& \qquad   -  2 \frac{1}{n^2 \h^2} \sumi \sumj \left\{ \l^0_{\i}(X_i)^2 \br_{\tilde{p}}(X_{d_n,i})' \left( \bm{\tilde{G}}_+^{-1} + \bm{\tilde{G}}_-^{-1} \right) ((K_+ + K_-) \br_{\tilde{p}})(X_{d_n,i}) \e_i \e_j \right\}    		\\
& \qquad   +  \frac{1}{n^3 \h^3} \sumi \sumj \sumk \left\{ \l^0_{\i}(X_i)^2 \br_{\tilde{p}}(X_{d_n,i})'  \left( \bm{\tilde{G}}_+^{-1} + \bm{\tilde{G}}_-^{-1} \right) ((K_+ + K_-) \br_{\tilde{p}})(X_{d_n,i}) \e_j \e_k  \right\} ,   		\\
W_{\i,2} & =  \frac{1}{n\h}\sumi \left\{  \l^0_{\i}(X_i)^2  v(X_i)^2  - \E[ \l^0_{\i}(X_i)^2  v(X_i)^2] \right\}  +  2 \frac{1}{n^2 \h^2} \sumi\sumj \l_{\i}^2(X_i, X_j) \l_{\i}^0(X_i) v(X_i),   		\\
W_{\i,3} & = \frac{1}{n^3 \h^3} \sumi \sumj \sumk \l_{\i}^1(X_i, X_j) \l_{\i}^1(X_i,X_k) v(X_i)   +   2 \frac{1}{n^3 \h^3} \sumi \sumj \sumk \l_{\i}^2(X_i, X_j, X_k) \l_{\i}^0(X_i) v(X_i),   			\\
B_{\i,1} & = \tO \frac{1}{n\h} \sumi \l^0_{\i}(X_i) \left([\mu_+(X_i) - \br_{\tilde{p}}(X_i - x)' \bbeta_{+,\tilde{p}}]   -  [\mu_-(X_i) - \br_{\tilde{p}}(X_i - x)' \bbeta_{-,\tilde{p}}]\right),  		\\
E_{\i,1} & = \tO  \frac{1}{n\h} \sumi \l^0_{\i}(X_i) \e_i,  		\\
E_{\i,2} & = \tO  \frac{1}{(n\h)^2} \sumi \sumj \l^1_{\i}(X_i, X_j) \e_i,  		\\
E_{\i,3} & = \tO  \frac{1}{(n\h)^3} \sumi \sumj \sumk \l^2_{\i}(X_i, X_j, X_k) \e_i,
\end{align*}
with the final line defining $\l^2_{\i}(X_i, X_j, X_k)$ in the obvious way following $\l^1_{\i}$, i.e.\ taking account of the next set of remainders. Terms involving $\l^2_{\i}(X_i, X_j, X_k)$ are higher-order, which is why $\l^2_{\i}$ is not needed in Section \ref{supp:terms}.

Straightforward moment calculations yield, where ``$\E[\ti] \oeq$'' denotes moments of the Taylor expansion above,
\begin{align*}
\E[\ti] & \oeq \st_{\i}^{-1} \E\left[B_{\i,1}\right]  - \frac{1}{2 \st_{\i}^2} \E\left[ W_{\i,1} E_{\i,1}\right],
\end{align*}
\begin{align*}
\E[\ti^2] & \oeq  \frac{1}{\st_{\i}^2} \E\left[ E_{\i,1}^2  +  E_{\i,2}^2 + 2 E_{\i,1} E_{\i,2}  + 2 E_{\i,1} E_{\i,3}  \right]   		\\
& \quad - \frac{1}{\st_{\i}^4} \E \left[ W_{\i,1} E_{\i,1}^2  +  W_{\i,2} E_{\i,1}^2  +   W_{\i,3} E_{\i,1}^2   + 2 W_{\i,2} E_{\i,1} E_{\i,2}   \right]  		\\
& \quad + \frac{1}{\st_{\i}^6} \E\left[ W_{\i,1}^2 E_{\i,1}^2  +  W_{\i,2}^2 E_{\i,1}^2 \right]  +  \frac{1}{\st_{\i}^2} \E\left[ B_{\i,1}^2 \right] - \frac{1}{\st_{\i}^4} \E\left[ W_{\i,1} E_{\i,1} B_{\i,1} \right],
\end{align*}
\begin{align*}
\E[\ti^3] & \oeq  \frac{1}{ \st_{\i}^3} \E\left[ E_{\i,1}^3 \right]    -   \frac{3}{2\st_{\i}^5} \E \left[ W_{\i,1} E_{\i,1}^3  \right]  	+   \frac{3}{\st_{\i}^3} \E\left[  E_{\i,1}^2 B_{\i,1} \right],
\end{align*}
and
\begin{align*}
\E[\ti^4] & \oeq  \frac{1}{\st_{\i}^4} \E\left[ E_{\i,1}^4  +  4 E_{\i,1}^3 E_{\i,2}  +  4 E_{\i,1}^3 E_{\i,3}  + 6 E_{\i,1}^2 E_{\i,3}^2  \right]   		\\
& \quad - \frac{2}{\st_{\i}^6} \E \left[ W_{\i,1} E_{\i,1}^4  +  W_{\i,2} E_{\i,1}^4  +   4 W_{\i,2} E_{\i,1}^3 E_{\i,2}  + W_{\i,3} E_{\i,1}  \right]  		\\
& \quad +  \frac{3}{\st_{\i}^8} \E \left[ W_{\i,1}^2 E_{\i,1}^4  +  W_{\i,2}^2 E_{\i,1}^4   \right]  		\\
& \quad + \frac{4}{\st_{\i}^4} \E\left[ E_{\i,1}^3  B_{\i,1} \right]   -   \frac{8}{\st_{\i}^6} \E\left[ W_{\i,1} E_{\i,1}^3 B_{\i,1} \right]  +  \frac{6}{\st_{\i}^4} \E\left[ E_{\i,1}^2 B_{\i,1}^2 \right].
\end{align*}
Computing each factor, we get the following results. For these terms below, indexes $i$, $j$, and $k$ are always distinct (i.e.\ $X_{\h,i} \neq X_{\h,j} \neq X_{\h,k}$). First, $\E\left[B_{\i,1}\right]$ is simply the fixed-$n$ version of the bias terms.
\begin{align*}
\E\left[ W_{\i,1} E_{\i,1}\right] & \oeq \tO^{-1} \E\left[ \h^{-1} \l_{\i}^0(X_i)^3  \e_i^3 \right],   		\\
\E\left[  E_{\i,1}^2 \right] & \oeq \st_{\i}^2,   		\\
\E\left[  E_{\i,1}E_{\i,2} \right] & \oeq \tO^{-2}  \E\left[ \h^{-1} \l^1_{\i}(X_i,X_i) \l^0_{\i}(X_i) \e_i^2 \right],   		\\ 
\E\left[ E_{\i,2}^2 \right] & \oeq \tO^{-1} \E\left[ \h^{-2} \l^1_{\i}(X_i, X_j)^2 \e_i^2 \right],   		\\
\E\left[  E_{\i,2}E_{\i,3} \right] & \oeq \tO^{-2}  \E\left[ \h^{-2} \l_v^2(X_i, X_j, X_j) \l^0_{\i}(X_i) \e_i^2 \right],   		\\ 
\E\left[  W_{\i,1}E_{\i,1}^2 \right] & \oeq \tO^{-2} \Biggl\{  \E\left[ \h^{-1} \l^0_{\i}(X_i)^4 \left( \e_i^4 - v(X_i)^2\right) \right]   		\\
& \quad  -  2 \st_{\i}^2 \E\left[ \h^{-1} \l^0_{\i}(X_i)^2 \br_{\tilde{p}}(X_{d_n,i})'  \left( \bm{\tilde{G}}_+^{-1} + \bm{\tilde{G}}_-^{-1} \right)\bm{\tilde{G}}^{-1} ((K_+ + K_-) \br_{\tilde{p}})(X_{d_n,i}) \e_i^2 \right]   		\\ 
& \quad  -  4  \E\left[ \h^{-1} \l^0_{\i}(X_i)^4 \br_{\tilde{p}}(X_{d_n,i})' \left( \bm{\tilde{G}}_+^{-1} + \bm{\tilde{G}}_-^{-1} \right)\e_i^2  \right] \E\left[ \h^{-1}  ((K_+ + K_-) \br_{\tilde{p}})(X_{d_n,i}) \l^0_{\i}(X_i) \e_i^2 \right]   		\\ 
& \quad  + \st_{\i}^2 \E\left[ \h^{-2} \l^0_{\i}(X_i)^2 \left(  \br_{\tilde{p}}(X_{d_n,i})' \left( \bm{\tilde{G}}_+^{-1} + \bm{\tilde{G}}_-^{-1} \right) ((K_+ + K_-) \br_{\tilde{p}})(X_{d_n,j}) \right)^2 \e_j^2 \right]   		\\ 
& \quad +  2 \E \bigg[ \h^{-1} \l^0_{\i} (X_j)^2 \Big( \E \Big[ \h^{-1} \br_{\tilde{p}}(X_{d_n,j})'  \left( \bm{\tilde{G}}_+^{-1} + \bm{\tilde{G}}_-^{-1} \right)    			\\
& \qquad \qquad \qquad\qquad\qquad   \times   ( (K_+ + K_-) \br_{\tilde{p}})(X_{d_n,i}) \l^0_{\i}(X_i)  \e_i^2 \vert X_j \Big] \Big)^2 \bigg]    \Biggr\},   		\\
\E\left[  W_{\i,2}E_{\i,1}^2 \right] & \oeq \tO^{-2} \Bigl\{ \E \left[ \h^{-1} \left( \l^0_{\i}(X_i)^2 v(X_i) - \E[\l^0_{\i}(X_i)^2 v(X_i)] \right) \l^0_{\i}(X_i)^2 \e_i^2 \right]     		\\
& \quad + 2 \st_{\i}^2 \E \left[ \h^{-1} \l^1_{\i}(X_i, X_i) \l^0_{\i}(X_i) v(X_i) \right]  \Bigr\},   		\\
\E\left[  W_{\i,2}E_{\i,1}E_{\i,2} \right] & \oeq \tO^{-2} \Bigl\{ \E \left[ \h^{-2} \left( \l^0_{\i}(X_j)^2 v(X_j) - \E[\l^0_{\i}(X_j)^2 v(X_j)] \right)  \l^1_{\i}(X_i, X_j) \l^0_{\i}(X_i) \e_i^2 \right]     		\\
& \quad + 2  \E \left[ \h^{-3} \l^1_{\i}(X_i, X_j)  \l^1_{\i}(X_k, X_j) \l^0_{\i}(X_i) \l^0_{\i}(X_k) v(X_i) \e_k^2 \right]  \Bigr\},   		\\
\E\left[  W_{\i,3}E_{\i,1}^2 \right] & \oeq \tO^{-2} \Bigl\{ \st_{\i}^2 \E \left[ \h^{-2} \left( \l^1_{\i}(X_i, X_j)^2  + 2\l^2_{\i}(X_i, X_j, X_j) \right) v(X_i)  \right]  \Bigr\},   		\\
\E\left[  W_{\i,1}^2 E_{\i,1}^2 \right] & \oeq \tO^{-2} \Bigl\{ \st_{\i}^2 \E \left[ \h^{-1}  \l^0_{\i}(X_i)^4 \left( \e_i^4 - v(X_i)^2 \right) \right]  + 2 \E\left[ \h^{-1} \l^0_{\i}(X_i)^3   \e_i^3  \right]^2  \Bigr\},   		\\
\E\left[  W_{\i,2}^2 E_{\i,1}^2 \right] & \oeq \tO^{-2} \st_{\i}^2 \Bigl\{ \E \left[ \h^{-1}  \left( \l^0_{\i}(X_i)^2 v(X_i)  -  \E[\l^0_{\i}(X_i)^2 v(X_i)] \right)^2 \right]  		\\
& \quad  + 4 \E\left[ \h^{-2}  \left( \l^0_{\i}(X_i)^2 v(X_i)  -  \E[\l^0_{\i}(X_i)^2 v(X_i)] \right) \l^1_{\i}(X_j, X_i) \l^0_{\i}(X_j) v(X_j) \right]   		\\
& \quad  + 4 \E\left[ \h^{-3} \l^1_{\i}(X_i, X_j) \l^0_{\i}(X_i) v(X_i) \l^1_{\i}(X_k, X_j) \l^0_{\i}(X_k) v(X_k) \right] \Bigr\},   		\\
\E\left[  W_{\i,1}E_{\i,1}B_{\i,1}\right]  & \oeq  \E\left[  W_{\i,1}E_{\i,1} \right] \E\left[B_{\i,1}\right],   		\\
\E\left[  E_{\i,1}^3 \right] & \oeq \tO^{-1} \E\left[ \h^{-1} \l^0_{\i}(X_i)^3 \e_i^3 \right],   		\\
\E\left[  W_{\i,1} E_{\i,1}^3 \right] & \oeq \E\left[  E_{\i,1}^2 \right]  \E\left[  W_{\i,1} E_{\i,1} \right],   		\\
\E\left[  E_{\i,1}^4 \right] & \oeq 3 \st_{\i}^4  +  \tO^{-2} \E\left[ \h^{-1} \l^0_{\i}(X_i)^4 \e_i^3 \right],   		\\
\E\left[  E_{\i,1}^3 E_{\i,2} \right] & \oeq  \tO^{-2} 6 \st_{\i}^2 \E\left[ \h^{-1} \l^1_{\i}(X_i, X_i) \l^0_{\i}(X_i) \e_i^2 \right],   		\\
\E\left[  E_{\i,1}^3 E_{\i,3} \right] & \oeq  \tO^{-2} 3 \st_{\i}^2 \E\left[ \h^{-2} \l^2_{\i}(X_i, X_j, X_j) \l^0_{\i}(X_i) \e_i^2 \right],   		\\
\E\left[  E_{\i,1}^2 E_{\i,2}^2 \right] & \oeq  \tO^{-2} \Bigl\{ \st_{\i}^2 \E\left[ \h^{-2} \l^1_{\i}(X_i, X_j)^2 \e_i^2 \right]   +   2  \E \left[ \h^{-3} \l^1_{\i}(X_i, X_j)  \l^1_{\i}(X_k, X_j) \l^0_{\i}(X_i) \l^0_{\i}(X_k) \e_i^2 \e_k^2 \right]  \Bigr\},   		\\
\E\left[  W_{\i,1} E_{\i,1}^4 \right] & \oeq \tO^{-2} \Bigl\{ \E\left[ \h^{-1} \l^0_{\i}(X_i)^3 \e_i^3 \right] \E\left[ \h^{-1} \l^0_{\i}(X_i)^3 \e_i^3 \right]  +  6 \E\left[  E_{\i,1}^2 \right]  \E\left[  W_{\i,1} E_{\i,1}^2 \right]  \Bigr\},   		\\
\E\left[  W_{\i,2} E_{\i,1}^4 \right] & \oeq \tO^{-2}\st_{\i}^2  6 \Bigl\{ \E \left[ \h^{-1} \left( \l^0_{\i}(X_i)^2 v(X_i) - \E[\l^0_{\i}(X_i)^2 v(X_i)] \right) \l^0_{\i}(X_i)^2 \e_i^2 \right]     		\\
& \quad + 2  \E \left[ \h^{-2} \l^1_{\i}(X_i, X_j) \l^0_{\i}(X_i) \l^0_{\i}(X_j)^2 \e_j^2 v(X_i) \right]   +  \E\left[ \h^{-1} \l^1_{\i}(X_i, X_i) \l^0_{\i}(X_i) v(X_i) \right] \Bigr\},   		\\
\E\left[  W_{\i,2} E_{\i,1}^3 E_{\i,2} \right] & \oeq 3 \E\left[   E_{\i,1}^2 \right] \E\left[  W_{\i,2} E_{\i,1} E_{\i,2} \right] ,  		\\
\E\left[  W_{\i,3} E_{\i,1}^4 \right] & \oeq 3 \E\left[   E_{\i,1}^2 \right] \E\left[  W_{\i,3} E_{\i,1}^2 \right] ,  		\\
\E\left[  W_{\i,1}^2 E_{\i,1}^4 \right] & \oeq 3 \E\left[   E_{\i,1}^2 \right] \E\left[  W_{\i,1}^2 E_{\i,1}^2 \right] ,  		\\
\E\left[  W_{\i,2}^2 E_{\i,1}^4 \right] & \oeq 3 \E\left[   E_{\i,1}^2 \right] \E\left[  W_{\i,2}^2 E_{\i,1}^2 \right] .
\end{align*}

The so-called approximate cumulants of $\ti$, denoted here by $\kappa_{\i,k}$ for the $k^{\text{th}}$ cumulant, can now be directly calculated from these approximate moments using standard formulas, such as Equation (2.6) of \citet*{Hall1992_book} which then become the terms of the expansion. See \citet*{Hall1992_book} for the general case and \citet*{Calonico-Cattaneo-Farrell_2018_JASA,Calonico-Cattaneo-Farrell_2019_CEOptimal} in the context of nonparametric regression.

\section{Details of Practical Implementation}
\label{supp:practical}

We now give details on practical issues that are discussed in the main text. These include the direct plug-in (DPI) rule to implement the coverage-error optimal bandwidth, variance estimation (bias estimation is discussed in Section \ref{supp:bias}), and the optimal choices $\rho^*$. These methods are implemented in {\sf R} and {\tt STATA} via the {\tt rdrobust} package, available from \url{http://sites.google.com/site/rdpackages/rdrobust}.

\subsection{Bandwidth Choice: Direct Plug-In (DPI)}
\label{supp:bandwidth}

In order to implement the plug-in bandwidth $\hat{\h}_{\RBC}$, we always set $K=L$ and $q=p+1$. The main steps are:
\begin{enumerate}[label=(\arabic*)]
	
	\item As a pilot bandwidth, use $\hat{h}_\MSE$: any data-driven version of $\h_\MSE$.
	
	\item Using this bandwidth, estimate $\bhat_{+,q}$ and $\bhat_{-,q}$ on each side of the threshold. 
	Then, form $\hat{\e}_{+,i} =  Y_i - \br_q(X_i - \x)'\bhat_{+,q}$ and $\hat{\e}_{-,i} = Y_i - \br_q(X_i - \x)'\bhat_{-,q}$.
	
	\item Using the pilot bandwidth and a choice of $\rho$, estimate the terms $\mathscr{Q}_{\RBC,k}$, $k=1,2,3$. As discussed more just below, from the formulas in Section \ref{supp:terms}, the estimates are defined by replacing:
	\begin{enumerate}[label=(\roman*)]
		\item $\h$ with $\hat{h}_\MSE$,
		\item population expectations with sample averages,
		\item residuals $\e_i$ with $\hat{\e}_i$, and
		\item limiting matrices with the corresponding sample versions using the pilot bandwidth. 
	\end{enumerate}			
	
	\item To estimate the bias constants $\tilde{\mathscr{B}}_\BC$, we follow \citet[Section 4.2]{Fan-Gijbels_1996_Book} and estimate derivatives $\mu^{(p+2)}$ using a global least squares polynomial fit of order $p+4$ on each side of the threshold.
	
	\item Finally we obtain:
	\[\hat{\h}_\RBC = \hat{\mathscr{H}} \; n^{-1/(3+p)}, \qquad
	\hat{\mathscr{H}} = \argmin_{H>0} \left\vert \frac{1}{H} \hat{\mathscr{Q}}_{\RBC,1}
	+ H^{5+2p} \hat{\mathscr{Q}}_{\RBC,2}
	+ H^{2+p} \hat{\mathscr{Q}}_{\RBC,3}\right\vert,\]

\end{enumerate}

Consistency of this bandwidth, meaning $\hat{\h}_\RBC / \h_\RBC \to_\P 1$, will follow under natural conditions. In particular, all that is required is consistent estimates for the constants appearing in $\mathscr{Q}_{\RBC,k}$, $k=1,2,3$, as listed in Section \ref{supp:terms}. The constants involved are fixed-$n$ computations, and so by ``consistent'' we mean $\hat{\mathscr{Q}}_{\RBC,1}/\mathscr{Q}_{\RBC,k} \to_\P 1$. All of the constants involved are kernel-weighted population averages, which may or may not involve $\mu_+(x)$ and  $\mu_-(x)$ or their derivatives. Using pilot bandwidths these can be consistently estimated by sample analogues. 

For example, the obvious estimator of $\Gpt{-}(\h) = \E[\h^{-1} (K_- \br_p \br_p')(\Xhi)]$ is, for some pilot bandwidth $\bar{\h}$, $\Gp{-}(\bar{\h}) = \sumi \big(K_- \br_p \br_p'\big)\big((X_i - \x)/\bar{\h}\big)/ n\bar{\h}$. If $n \bar{\h} \to \infty$, a law of large numbers yields that $\Gp{-}(\bar{\h})$ is consistent for its fixed-$n$ expectation, as in $\Gp{-}(\bar{\h})/\E[\Gp{-}(\bar{\h})] \to_\P 1$. If $\h \vee \bar{\h} \to 0$ then the limits of both fixed-$n$ expectations agree, $\E[\Gp{-}(\bar{\h})]/\Gpt{-}(\h) \to 1$. This yields the desired result.

The logic for all the remaining terms is similar, with the possible addition of a consistent estimator for $\mu_+$ or $\mu_-$, and the associated estimated residuals, variances, and biases. These are also easily formed based on pilot bandwidths, for example using rule-of-thumb implementations of the respective MSE-optimal choice for the specific problem. As an example, consider estimating $\mathscr{Q}_{\RBC,3} = 2 \phi(z_{\alpha/2}) Q_{\RBC,3}(z_{\alpha/2}) \tilde{\mathscr{B}}_\BC$. This requires estimates of $Q_{\RBC,3}(z_{\alpha/2})$ and $\tilde{\mathscr{B}}_\BC$. The former term is $Q_{\RBC,3}(z)  =   \st_{\i}^{-4} \E [ \h^{-1} \l^0_{\i}(X_i)^3 \e_i^3 ]  \left\{ z^3 / 3 \right\}$. First, $\st_{\i}^{-4}$ can be estimated by employing $\shat_\ti^2$ following Section \ref{supp:variance}: all that is required is a pilot bandwidth that delivers consistent estimates of $\mu_+$ and $\mu_-$, for which any ROT MSE choice will do, and estimates of other sample averages, which follow as above and can use the same pilot bandwidth. Notice that $\st_{\i}^2  = \E[ \h^{-1} \l^0_{\i}(X)^2 v(X) ]$, and so if we can estimate this quantity it is obvious that replacing the squaring with cubing estimates the factor $\E [ \h^{-1} \l^0_{\i}(X_i)^3 \e_i^3 ]$, and altogether we find that $\hat{Q}_{\RBC,3}(z_{\alpha/2})(\bar{\h}) / Q_{\RBC,3}(z_{\alpha/2})(\h) \to_\P 1$. Estimation of the bias term follows the same way, and we follow \citet[Section 4.2]{Fan-Gijbels_1996_Book}.

\subsection{Alternative Standard Errors}
\label{supp:standard errors}

We consider two alternative estimates of $\bS_+$ and $\bS_-$ than those presented in Section \ref{supp:variance}.  First, motivated by the fact that the least-squares residuals are on average too small, we propose HC$k$ heteroskedasticity-consistent estimators; see \citet*{MacKinnon2013_BookChap} for details and a recent review. \citet*{Calonico-Cattaneo-Farrell-Titiunik_2019_RESTAT} discuss how they can be applied in the context of local polynomial estimation to construct $\shat_\i^2$-HC$k$, $k=0,1,2,3$, where $\shat_\i^2$-HC0 is the original estimator presented above and the others use different weights based on projection matrices.

A second option is to use a nearest-neighbor-based variance estimators with a fixed number of neighbors, following the ideas of \citet*{Muller-Stadtmuller1987_AoS} and \citet*{Abadie-Imbens2008_AdES}. To define these, let $J$ be a fixed number and $j(i)$ be the $j$-th closest observation to $X_i$, $j=1, \ldots, J$, and set $ \hat{\e}_{+,i} = \I(X_i\geq c)\sqrt{\frac{J}{J+1}} ( Y_i  -  \sum_{j=1}^J Y_{j(i)} / J )$, $ \hat{\e}_{-,i} = \I(X_i < c)\sqrt{\frac{J}{J+1}} ( Y_i  -  \sum_{j=1}^J Y_{j(i)} / J )$.

As discussed in \citet*{Calonico-Cattaneo-Farrell_2018_JASA}, both types of residual estimators could be handled in our results under natural modifications.

\subsection{Equivalent Kernels}

We discuss how to optimize the asymptotic variance constant featuring the length of the RBC confidence interval estimator using the \emph{equivalent kernel representation} of local polynomials; see Section 3.2.2 of \citet*{Fan-Gijbels_1996_Book}. Detailed derivations are found there.

For simplicity, consider the one-sided bias-corrected estimate of $\mu_{+}$, i.e., half of $\that_{0,\BC} = \that_0 - \h^{p+1}  \hat{\mathscr{B}}$. The same of course holds for the ``$-$'' half of $\that_{0,\BC}$. Recall the definitions in and around \eqref{suppeqn:that rbc} and that $q=p+1$. Then we consider
\begin{align*}
\hat{\mu}_{+,\BC}^{(0)}(\x) = \hat{\mu}_{+,\BC}  = \frac{1}{n} \be_0'\Gp{+}^{-1} \Orbc{+}  \bY  &	= \frac{1}{n} \be_0'\Gp{+}^{-1} \left( \Op{+} - \rho^{p+1} \Lp{+} \be_{p+1}' \Gq{+}^{-1} \Oq{+} \right)  \bY   		\\
& =: \frac{1}{n\h} \sumi \mathcal{K}^\BC_{+,p}\big(\Xhi; K, \rho \big) Y_i,
\end{align*}
where the last equality defines the weights (recall the definitions of $\Op{+}$ and $\Oq{+}$)
\[ \mathcal{K}^\BC_{+,p}\big(x; K, \rho \big) = \be_0'\Gp{+}^{-1} \Big[  (K_+ \br_p)(x) - \rho^{p+2} \Lp{+} \be_{p+1}' \Gq{+}^{-1}  (K_+ \br_q)(\rho x) \Big]. \]
This function depends on the sample through $\Gp{+}$, $\Lp{+}$, and $\Gq{+}$. To find the equivalent kernel, we replace these with their limiting versions. Note that here, as opposed to elsewhere in the paper, we use the population limiting versions, not fixed-$n$ expectations, i.e.\ we need the \emph{limit} of $\Gpt{+} = \E[\Gp{+}]$. Under our assumptions, $\Gp{+} \to_\P f(\x)\Gpbar{+}$, $\Lp{+} \to_\P f(\x)\Lpbar{+}$, and $\Gq{+}^{-1} \to_\P f(\x)\Gqbar{+}$, at sufficient fast rates, such that
\begin{align*}
\hat{\mu}_{+,\BC} & = \frac{1}{n\h}\sumi \bar{\mathcal{K}}^\BC_{+,p}\big(\Xhi; K, \rho \big) Y_i \; \{1 + o_\P(1)\},
\end{align*}
where the equivalent kernel is
\[  \bar{\mathcal{K}}^\BC_{+,p}\big(x; K, \rho \big) =  \frac{1}{f(\x)} \be_0'\Gpbar{+}^{-1} \Big[ (K_+ \br_p)(x) - \rho^{p+2} \Lpbar{+} \be_{p+1}' \Gqbar{+}^{-1} (K_+ \br_q)(\rho x) \Big], \]
with \[ \Gpbar{+} = \int (K_+ \br_p\br_p')(u)du,    \qquad     \Lpbar{+} = \int K_+(u) \br_p(u)u^{p+1}du    \qquad \text{and} \qquad     \Gqbar{+} = \int (K_+ \br_q\br_q')(u)du. \]

The shape of this equivalent kernel depends on the initial kernel chosen, $K(\cdot)$, and $\rho$. \citet*{Cheng-Fan-Marron_1997_AoS} show that the asymptotic variance of a local polynomial point estimator at a boundary point is minimized by employing the uniform kernel $K(u)=\I(|u|\leq 1)$. The resultant equivalent kernel (the ``optimal'' equivalent kernel) will be denoted $\mathcal{K}^*_{+,p}(x)$ for any $p$. If the uniform kernel is used when forming $\irbc(\h)$, then $\rho=1$ is optimal in terms of minimizing the asymptotic constant featuring the interval length: that is, $\rho=1$ makes the induced equivalent kernel, $\bar{\mathcal{K}}^\BC_{+,p}\big(x; K, \rho \big)$, pointwise equal to the optimal equivalent kernel, $\mathcal{K}^*_{+,p+1}(x)$.

\begin{figure}[h]
	\centering
	\caption{$\mathcal{K}^*_{+,p+1}(x)$ vs. $\bar{\mathcal{K}}^\BC_{+,p}\big(x; K, \rho^* \big)$}\label{fig:rho}
	
	\begin{subfigure}[b]{0.5\textwidth}
		\includegraphics[width=1.05\textwidth]{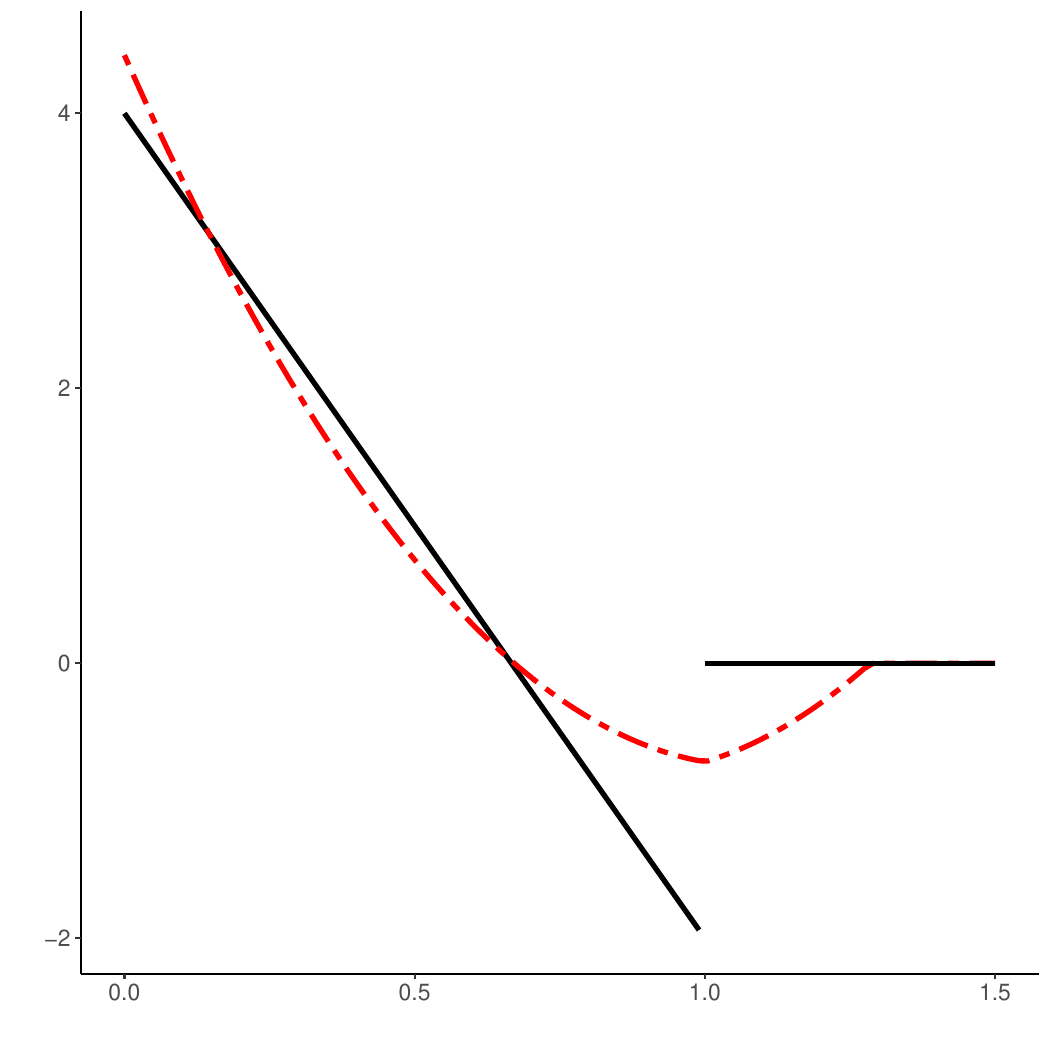}
		\caption{$p=0$}
	\end{subfigure}%
	\begin{subfigure}[b]{0.5\textwidth}
		\includegraphics[width=1.05\textwidth]{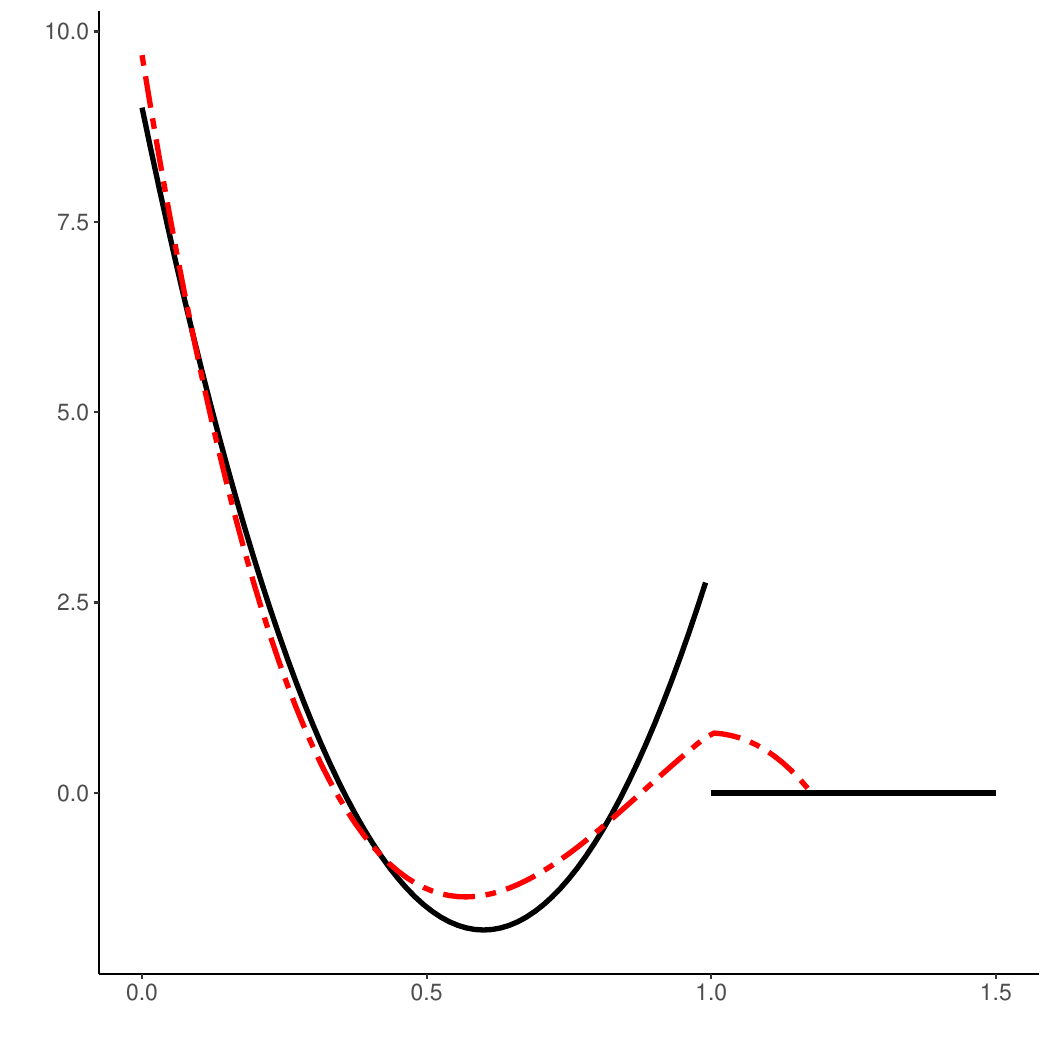}
		\caption{$p=1$}
	\end{subfigure}
	
	\begin{subfigure}[b]{0.5\textwidth}
		\includegraphics[width=1.05\textwidth]{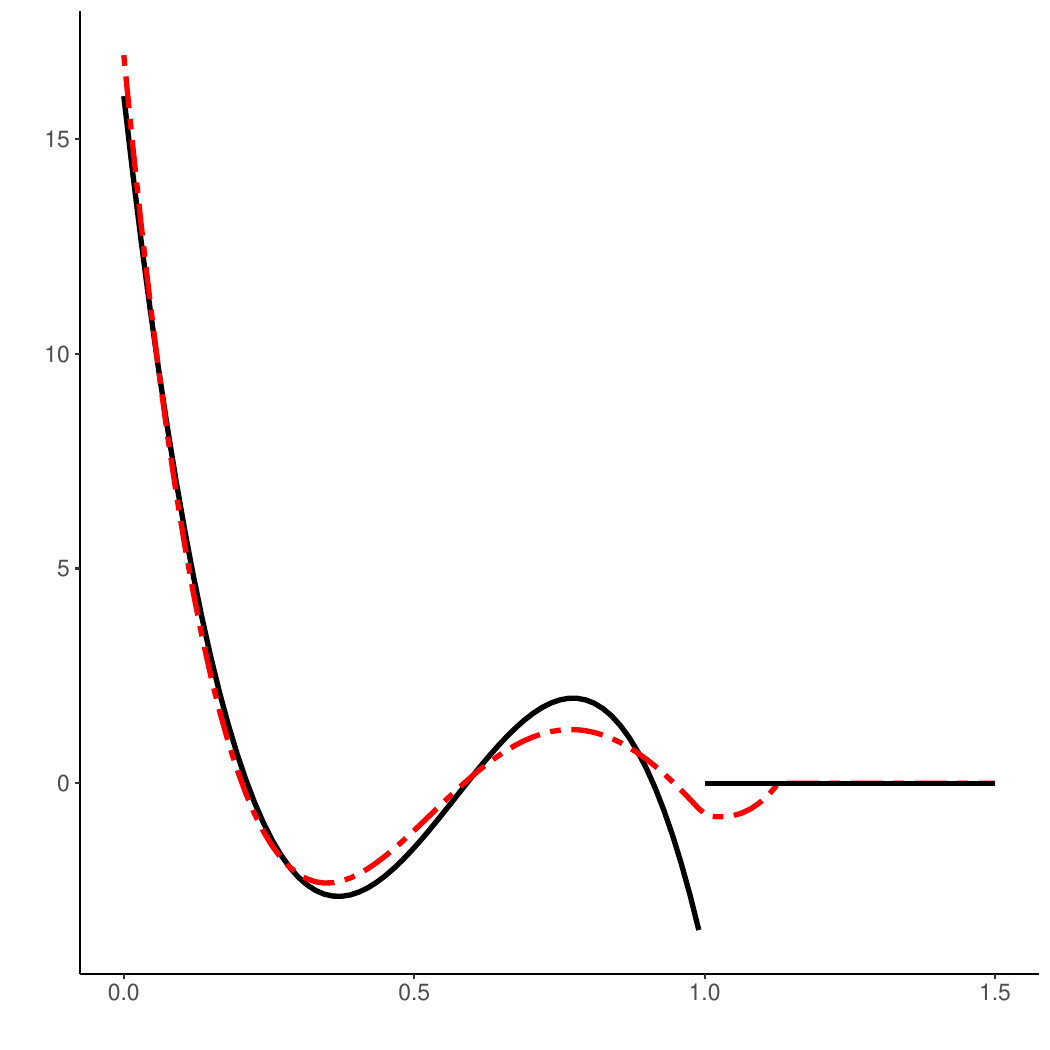}
		\caption{$p=2$}
	\end{subfigure}%
	\begin{subfigure}[b]{0.5\textwidth}
		\includegraphics[width=1.05\textwidth]{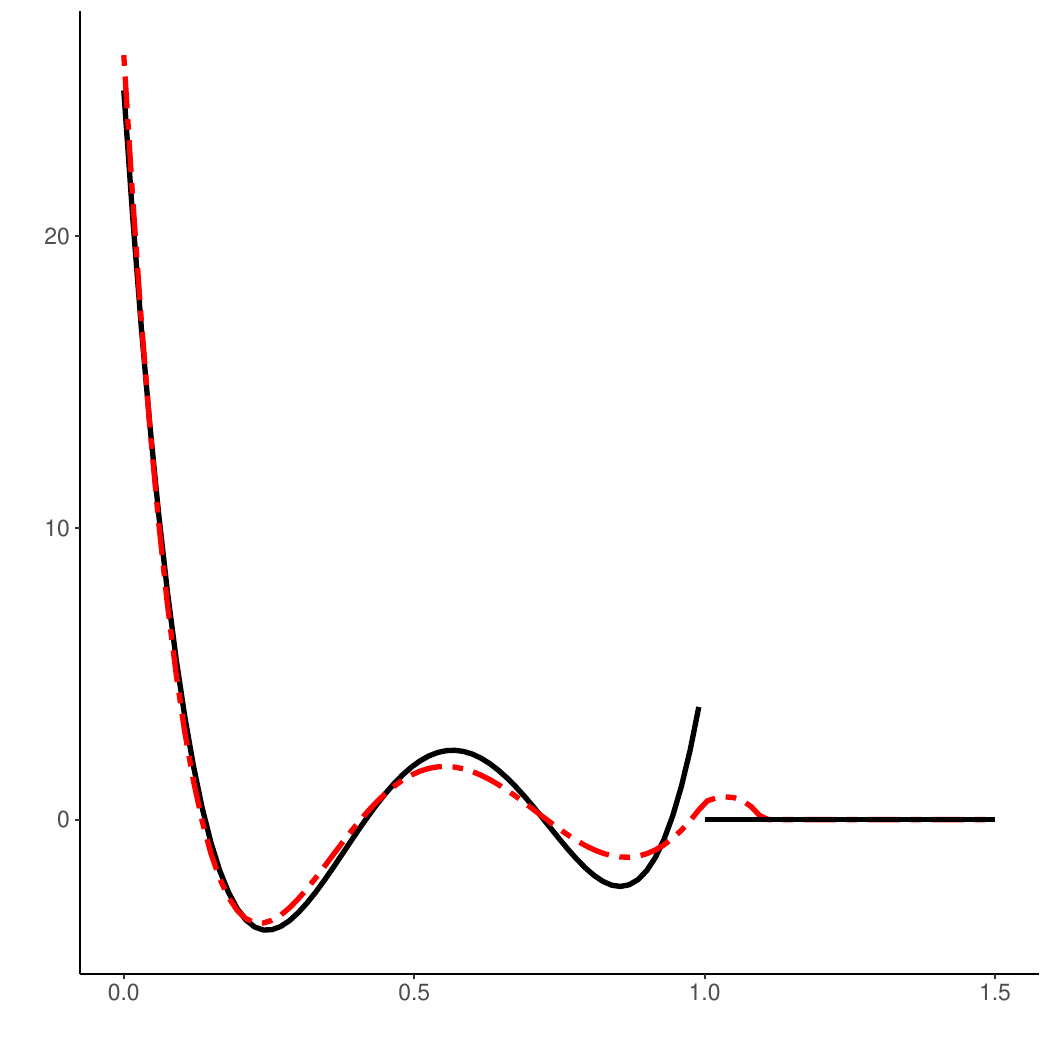}
		\caption{$p=3$}
	\end{subfigure}
	
\end{figure}

However, if a kernel other than uniform is used, we can find the optimal choice of $\rho$ in terms of minimizing the $L_2$ distance between the induced equivalent kernel, $\bar{\mathcal{K}}^\BC_{+,p}\big(x; K, \rho \big)$, and the optimal variance-minimizing equivalent kernel, $\mathcal{K}^*_{+,p+1}(x)$. To be precise, we compute
\[\rho^* = \argmin_{\rho>0} \int \left| \bar{\mathcal{K}}^\BC_{+,p}\big(x; K, \rho \big)  -  \mathcal{K}^*_{+,p+1}(x) \right| ^2 dx.\]
A common choice is the triangular kernel $K(u)=(1-|u|)\I(|u|\leq 1)$, which \citet*{Cheng-Fan-Marron_1997_AoS} show is MSE-optimal (i.e., optimal from a point estimation perspective). We illustrate the shape of the resulting equivalent kernel under the $L_2$-optimal choice of $\rho$ in Figure \ref{fig:rho} for the triangular bias-corrected equivalent kernel and different choices of $p$. The corresponding values of $\rho^*$ were given in Table 1 of the paper.

\clearpage

\clearpage

\bibliography{Calonico-Cattaneo-Farrell_2020_ECTJ--Bibliography}{}
\bibliographystyle{chicago}